\documentclass[twocolumn]{aastex6}
\usepackage{amsmath,amssymb}
\usepackage{hyperref,graphicx,subfigure} 
\usepackage{ot1patch}
\usepackage{gensymb}

\newcommand{\degrees}{^{\circ}}
\newcommand{\Msun}{\ensuremath{M_{\odot}}}

\newcommand{\bil}{1}
\newcommand{\itu}{2}
\newcommand{\sab}{3}
\newcommand{\fg}{4}

\begin{document}

\title{On the Magnetic Fields, Beaming Fractions, and Fastness Parameters of Pulsating Ultra-Luminous X-Ray Sources}

\author{M.H. Erkut\altaffilmark{\bil,\fg}, M.M. T\"urko{\u g}lu\altaffilmark{\itu}, K.Y. Ek\c{s}i\altaffilmark{\itu,\fg}, and M.A. Alpar\altaffilmark{\sab,\fg}}

\affil{\altaffilmark{\bil}Faculty of Engineering and Natural Sciences, Istanbul Bilgi University, 34060, Istanbul, Turkey\\
\altaffilmark{\itu}Physics Engineering Department, Faculty of Science and
Letters, Istanbul Technical University, 34469, Istanbul, Turkey\\
\altaffilmark{\sab}Faculty of Engineering and Natural Sciences,
Sabanc\i\ University, 34956, Istanbul, Turkey\\
\altaffilmark{\fg}Feza G\"{u}rsey Center for Physics and
Mathematics, Bo\u{g}azi\c{c}i University, 34684,
Istanbul, Turkey}

\begin{abstract}
The discovery of pulsating ultra-luminous X-ray sources (PULX) suggests that neutron stars are presumably common within the ultra-luminous X-ray source (ULX) population though the majority of the population members are currently lacking pulsations. These systems are likely to host neutron stars accreting mass at super-Eddington (super-critical) rates from their massive companion in high-mass X-ray binaries. Taking into account the spherization of the accretion flow in the super-critical regime, the beaming of X-ray emission, and the reduction of the scattering cross-section in a strong magnetic field, we infer the ranges for the neutron-star surface magnetic dipole field strengths, beaming fractions, and fastness parameters in the PULX M82~X-2, ULX~NGC~5907, ULX~NGC~7793~P13, NGC~300~ULX1, M51~ULX-7, NGC~1313~X-2, and Swift~J0243.6+6124 from a set of conditions based on a variety of combinations of different spin and luminosity states. Using the observed spin-up rates under the critical luminosity condition, we estimate the surface-field strengths in the $\sim 10^{11}-10^{13}\,{\rm G}$ range for all PULX. In general, the results of our analysis under the subcritical luminosity condition indicate surface-field strengths in the $\sim 10^{11}-10^{15}\,{\rm G}$ range. We argue that the PULX do not require magnetar-strength surface dipole fields if beaming is taken into account; yet the fields are strong enough for the neutron stars in ULX to magnetically channel the accretion flow in super-critical accretion disks. 
\end{abstract}


\section{Introduction}
\label{intro}

Each nearby galaxy hosts a few ultra-luminous X-ray sources (ULX) which are accreting compact objects whose luminosities well exceed the Eddington limit for a solar mass object.
The assumption that the luminosity is isotropic lead to the view that these systems host intermediate mass black holes \citep[IMBHs; see e.g.][]{col99,kon+04,mil+04,liu08} of mass $M_{\rm BH} \sim 100$--$10^4 \,\Msun$.
Many other studies argued that ULX host stellar mass black holes with
anisotropic radiation slightly exceeding the Eddington limit \citep[see e.g.][]{king+01,pou+07,gla+09}.
Yet another explanation was suggested by \citet{beg02} who considered that the Eddington limit is simply exceeded in these sources.

The discovery of pulsating ultra-luminous X-ray sources (PULX) in M82 \citep{bach+14}, NGC~5907 \citep{isr+16a}, NGC~7793 \citep{isr+16b,fue+16}, NGC 300 \citep{carp+18}, NGC~1313 \citep{sat+19}, M51 \citep{rod+20}, and our galaxy \citep{wil+18} implies that ULX hosting neutron stars, rather than black holes, are more common within the ULX population \citep{sha15,wik+17,pintore+17,mid17a,mid17b}.

The PULX represent the highest end of the luminosity distribution of accreting neutron stars that include the conventional X-ray pulsars in high-mass X-ray binaries \citep{nag89,bil+97}
and mildly super-Eddington X-ray pulsars \citep{tru+07,tru08,wen+17,tsy+17,tsy+18}. Initially,
\citet{bach+14} attributed the super-Eddington luminosity of the first PULX they discovered to fan beam geometry viewed at a favorable angle \citep{gne73} and inferred a magnetic field of $B\simeq 10^{12}\,{\rm G}$. Motivated by the rapid spin-up of the object, \citet{eks+15} suggested the neutron star in this system had super-strong magnetic fields ($B \gtrsim B_{\rm c} =4.4\times 10^{13}$~G) that reduce the scattering cross-section and increase the critical luminosity \citep{can71,pac92,mus+16} and speculated that the quadrupole fields could be even stronger, thus forming a link with the isolated magnetars in the galaxy. 
Later work \citep{dal+15,tsy+16} inferred magnetic fields of the same order. \citet{isr+16a} suggested that the extreme properties of the PULX in NGC~5907 can only be explained if super-strong quadrupole fields are invoked. \citet{kin19}, however, argued that the magnetar-strength fields are unlikely to form in binaries and ULX are no exception. They invoked magnetic fields in the $10^{11}- 10^{13}\,{\rm G}$ range and beaming of emission due to outflows and spherization of the disk in supercritical accretion \citep{sha73}.

Yet another line of reasoning presumes that the discovered PULX are away from spin equilibrium and infers much weaker magnetic fields $B \sim 10^{10}\,{\rm G}$ \citep{klu15,chr+18}. 
If PULX indeed have such small magnetic fields the magnetosphere enshrouded by a mass flux at the Eddington rate or higher would be too small (if not crushed to the surface) to allow any pulsations to be observed. Reducing the mass accretion rate by invoking beaming would also be problematic: what parameter leads to such strong beaming given both the magnetic field and mass flow rate are so ordinary? Comparison of the soft X-ray
emission to the observed He~\textsc{ii} $\lambda$4686 emission line luminosity
suggests that geometric beaming effects in the NGC 300 ULX-1 system do not involve beaming factors more than 5-6 \citep{bin+18}.
It is likely that both beaming and cross-section reduction play a role  \citep{ton15a,ton15b,mus+18} in addressing how PULX can exceed the Eddington limit expected from a $\sim 1$--$2\,\Msun$ object.

The 4.5 keV line discovered in the spectrum of the ULX in M51 \citep{bri+18} corresponds to a magnetic field of $10^{15}\,{\rm G}$ $(10^{11}\,{\rm G})$ if due to proton (electron) scattering. As the line is narrower than any electron cyclotron resonance spectral feature detected to date, \citet{bri+18} favor the proton interpretation \citep[but see][]{mid+19a}. Even stronger field estimates are obtained if the origin of the cyclotron line is not the surface of the star but the accretion column. The PULX NGC 300 ULX-1 has a potential cyclotron feature at $\sim 13\,{\rm keV}$, suggesting a magnetic field of $B \sim 10^{12}\,{\rm G}$ \citep{wal+18b}. These observations can be reconciled with the $B \sim 10^{13}\,{\rm G}$ surface fields inferred from strong spin-up only by assuming that they are electron cyclotron resonance spectral features originating from accretion column of height a few times the radius of the star.

According to the recently developed evolutionary model of ULX \citep{eea+19}, some newborn neutron stars with initial magnetic dipole fields in the $\sim 10^{11}$--$10^{15}\,{\rm G}$ range interact with wind-fed disks in high-mass X-ray binaries (HMXBs) to appear as nonpulsating and pulsating ultra-luminous sources at different stages of a lifetime of $\sim 10^6$~yr. In the early super-critical propeller phase, the rapidly rotating neutron-star magnetosphere transfers the spindown power to the accretion disk and the system turns out to be a nonpulsating ULX. It takes $\sim 10^5$~yr for systems with initial magnetic fields of $\sim 10^{11}\,{\rm G}$ to pass through the super-critical propeller phase appearing as ULX with luminosities typical of ultra-luminous super-soft sources. The strongly magnetized systems with initial field strengths in the $\sim 10^{13}$--$10^{15}\,{\rm G}$ range on the other hand, are highly likely to appear as PULX during the long accretion phase these systems must undergo in the course of their lifetimes \citep{eea+19}.

In this paper, we consider PULX as neutron stars accreting mass at super-Eddington (super-critical) rates from their massive normal companions in HMXBs. We infer the magnetic dipole field strengths on the neutron-star surface from the torque and luminosity states of PULX. The ranges for the beaming fractions and fastness parameters are also estimated. We speculate that many ULX host strongly magnetized neutron stars whose beamed X-ray luminosity coincides with our line of sight leading to the appearance of super-Eddington luminosities. This is supported by recent population synthesis results \citep{sha15} and evolutionary calculations \citep{fra+15}.

In Section~\ref{model}, we introduce the model equations describing the magnetosphere-disk interaction in the super-critical accretion regime. In Section~\ref{results_beaming}, we estimate the ranges for the magnetic field, beaming fraction, and fastness parameter of each PULX using information on the spin and luminosity states of the neutron star. We summarize our results and discuss their implications in Section~\ref{discuss}. Finally, we present our concluding remarks in Section~\ref{conc}.

\section{Model Equations} \label{model}

\subsection{Inner Radius} \label{sec:inner}

The accretion luminosity is given by $L_{\rm X}=GM_{\ast}\dot{M}_{\ast}/R_{\ast}=\epsilon \dot{M}_{\ast} c^2$,
where $\dot{M}_{\ast}$ is the mass accretion rate onto the neutron star of mass $M_{\ast}$ and radius $R_{\ast}$, $\epsilon\equiv GM_{\ast}/(R_{\ast}c^2)$ is the efficiency of gravitational energy release, and $c$ is the speed of light \citep{pri72}. Assuming that the X-ray emission is beamed by a factor $b<1$, the X-ray flux at a source distance $d$ is $F_{\rm X}=L_{\rm X}/ 4\pi b d^2$. Accordingly, the mass accretion rate
onto the neutron star can be estimated as
\begin{equation}
\dot{M}_{\ast} = \frac{4\pi b R_{\ast} d^2 F_{\rm X} }{ GM_{\ast} }.
\label{dotM}
\end{equation}
We assume the mass donor is at a stage such that it transfers matter at a super-Eddington rate, $\dot{M}_0>\dot{M}_{\rm E}\equiv L_{\rm E}/\epsilon c^2$,
where $L_{\rm E}=4{\rm \pi} G M_* m_{\rm p} c /\sigma_{\rm T}$ ($m_{\rm p}$ is the proton mass and $\sigma_{\rm T}$ is the Thomson cross-section of the electron) is the Eddington luminosity.

The super-critical mass flow within the disk causes the accretion flow to become quasi-spherical inside a critical radius (also known as the spherization radius)
\begin{equation}
R_{\rm sp} = \frac{27\epsilon \sigma_{\rm T} \dot{M}_0 }{8{\rm \pi} m_{\rm p} c} \simeq 1.43\times 10^9~{\rm cm}\,\, \epsilon \, \left( \frac{\dot{M}_0}{10^{20}\,{\rm g\,s^{-1}}} \right)
\end{equation}
determined by $L(R>R_{\rm sp})=27\epsilon GM_* \dot{M}_0/(2R_{\rm sp})=L_{\rm E}$ and the flow regulates itself so that the local Eddington rate is not exceeded, i.e., 
\begin{equation}
\dot{M} =
\begin{cases}
  \dot{M}_0(R/R_{\rm sp}), \quad & \mbox{for } R<R_{\rm sp}; \\
\dot{M}_0, \quad & \mbox{for } R>R_{\rm sp}.
\end{cases}
\label{mdot}
\end{equation}
\citep{sha73}. Within the spherization radius the disk is geometrically and optically thick. Axial symmetry is maintained while the flow has a dominant component in the spherical radial direction. Dynamical equations employed for the inflowing matter near the disk midplane are still $z$-averaged as for thin disks.

In the accretion regime, viscous stresses are negligible at the magnetopause and the inner radius of the disk is determined by the balance between magnetic and material stresses \citep{gho79a},
\begin{equation}
\frac{d}{dR}\left(\dot{M}R^2\Omega \right)=-R^2 B_{\phi}^{+} B_z, \label{amb}
\end{equation}
where $\Omega$ is the angular velocity of the innermost disk matter within the boundary region, $B_z$ is the poloidal magnetic field, $B_{\phi}^{+}=\gamma_{\phi} B_z$ is the toroidal magnetic field at the surface of the disk and $\gamma_{\phi}$ is the azimuthal pitch of order unity.

Integrating \autoref{amb} over the narrow zone (boundary region), where the neutron-star magnetosphere threads the inner disk, we write
\begin{equation}
\dot{M}_{\rm in} R_{\rm in}^2\left[\Omega_{\rm K}\left(R_{\rm in}\right)-\Omega_*\right]= -\int_{R_{\rm in}-\Delta R}^{R_{\rm in}} \gamma_{\phi} B_z^2 R^2 dR,
\label{stresses}
\end{equation}
approximating $\Omega\left(R_{\rm in}\right)$ and $\Omega\left(R_{\rm in}-\Delta R\right)$ with the Keplerian angular velocity at the inner radius, $\Omega_{\rm K}\left(R_{\rm in}\right)$, and the angular velocity of the neutron star, $\Omega_*$, respectively. Here, $\dot{M}_{\rm in}\equiv \dot{M}(R_{\rm in})$ is the mass flux at the innermost disk radius. Note that $\dot{M}_{\rm in}=\dot{M}_0\left(R_{\rm in}/R_{\rm sp}\right)$ for $R_{\rm in}<R_{\rm sp}$ and $\dot{M}_{\rm in}=\dot{M}_0$ for $R_{\rm in}>R_{\rm sp}$ in compliance with \autoref{mdot}. The integral on the right hand side of \autoref{stresses} is of the form $\mu^2\delta/R_{\rm in}^3$, where $\mu$ is the magnetic dipole moment of the neutron star and
\begin{equation}
\delta = \frac{\Delta R}{R_{\rm in}}
\label{delta}
\end{equation}
is the relative width of the coupled domain (boundary region) between the disk and the neutron-star magnetosphere. The azimuthal pitch can be expressed in terms of the rotational shear between the magnetosphere and the innermost disk matter as $\gamma_{\phi}\simeq \omega_{\ast}-1$, where
\begin{equation}
\omega_{\ast} \equiv \frac{\Omega_{\ast}}{\Omega_{\rm K}\left( R_{\rm in} \right)}=\left(\frac{R_{\rm in}}{R_{\rm co}}\right)^{3/2} \label{fstns}
\end{equation}
is the fastness parameter \citep{erk04}. Using the conservation of mass flux in the steady state, i.e., $\dot{M}_{\rm in}=\dot{M}_{\ast}$ and assuming that $B_z\simeq -\mu/R^3$, it follows from \autoref{stresses} and \autoref{dotM} that
\begin{equation}
R_{\rm in} = \left( \frac{\mu^2 \sqrt{GM_{\ast}}\delta}{4\pi b R_{\ast} d^2 F_{\rm X}}  \right)^{2/7}.
\label{R_in}
\end{equation}
Note that \autoref{R_in} accounts for both regimes of high mass accretion rates: super-critical disks with $R_{\rm in}<R_{\rm sp}$ and thin disks truncated by high magnetic fields with $R_{\rm in}>R_{\rm sp}$. This expression is, however, not explicit as it depends on $b$, which in turn might depend on the inner radius of the disk. In the next subsection, we determine $b$ in terms of $R_{\rm in}$ to obtain a self-consistent solution.

\subsection{Beaming Fraction}
\label{sec:beam}

We assume here that the beaming fraction, $b$,  is determined by the opening angle of the polar cap and therefore the fractional area of the polar region as
\begin{equation}
b = \frac{A_{\rm c}}{\gamma A},
\label{beam}
\end{equation}
where $A_{\rm c}$ is the polar cap area, $A=4\pi R_{\ast}^2$ is the total surface area of the neutron star, and $\gamma<1$ is a normalization constant corresponding to the maximum fractional area of the polar cap. The beaming definition of \citet{king09} depends only on the mass accretion rate (in units of Eddington mass accretion). The beaming, however, depends not only on the mass accretion rate but also on the magnetic field and the inclination angle as far as the pulsating sources are concerned. Considering the fact that most of the pulsations stem from magnetospheric accretion, we propose that the beaming defined in \autoref{beam} is more reliable for the PULX.

\begin{figure}
    \centering
            \includegraphics[width=\columnwidth]{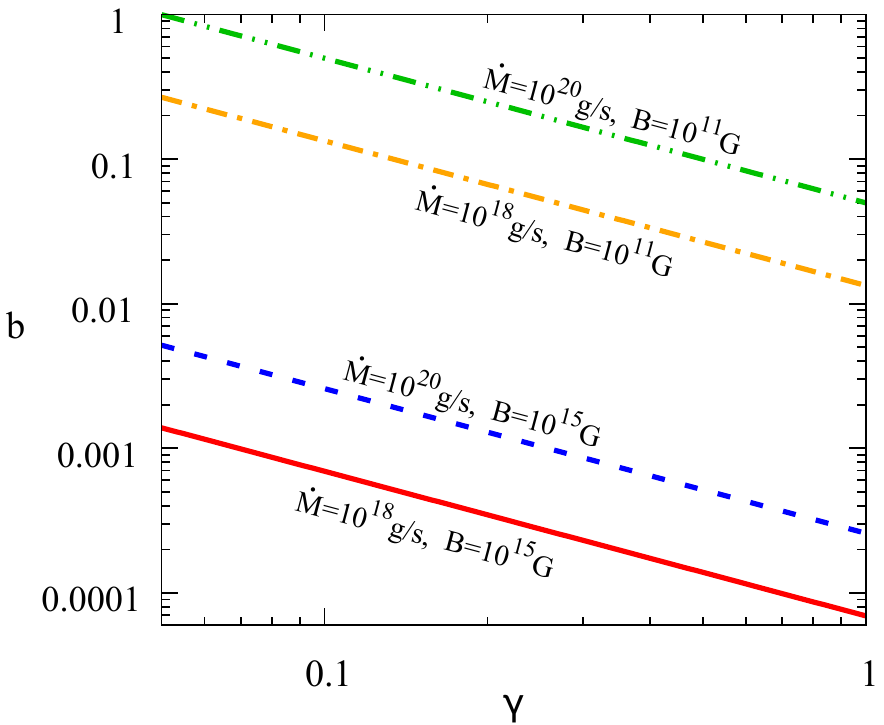}             
    \caption{Variation of the beaming fraction $b$ (vertical axis) with the normalization constant $\gamma$ (horizontal axis) for the $10^{18}$--$10^{20}\,{\rm g\,s^{-1}}$ range of $\dot{M}_*$ and $10^{11}$--$10^{15}\,{\rm G}$ range of $B$. For each line corresponding to the limits of the $\dot{M}_*$ and $B$ ranges, we choose the dimensionless width of the magnetic boundary region as $\delta=0.1$ and assume $M_*=1.4\,\Msun$, $R_*=10\,{\rm km}$, and $\alpha=20\degrees$ for the mass, radius, and inclination angle between the rotation and magnetic axes of the neutron star, respectively. The beaming fractions corresponding to the intermediate values of $\dot{M}_*$ and $B$ lie between the red solid line $(\dot{M}_*=10^{18}\,{\rm g\,s^{-1}},\,B=10^{15}\,{\rm G})$ and the green dot dot dashed line $(\dot{M}_*=10^{20}\,{\rm g\,s^{-1}},\,B=10^{11}\,{\rm G})$.}
   \label{bmng}
\end{figure}

The size of the polar cap can be inferred as follows:
Dipole field lines in polar coordinates are defined by $r=C\sin^2 \theta$. Assuming that the inclination angle between the rotation and magnetic axes is $\alpha$, the
field line that connects the inner edge of the disk to the star is described by
$r=R_{\rm in} \sin^2 \theta /\cos^2 \alpha$.
This field line makes an angle $\theta_{\rm c}$ with the magnetic axis on the neutron-star surface such that
$\sin \theta_{\rm c}=\sqrt{R_{\ast}/R_{\rm in}} \cos\alpha$. The ratio of the polar cap area to the surface area of the neutron star is $A_{\rm c}/A
=\frac12 (1-\cos \theta_{\rm c})$. For $R_{\ast} \ll R_{\rm in}$ ($\sin \theta_{\rm c} \simeq \theta_{\rm c}$ and $\cos \theta_{\rm c} \simeq 1 - \theta_{\rm c}^2/2$), as would be valid for a strongly magnetized neutron star, the same ratio yields
\begin{equation}
\frac{A_{\rm c}}{A} \simeq \frac{R_{\ast}}{4R_{\rm in}} \cos^2 \alpha \label{fracrat}
\end{equation}
\citep{fra+02}.
Substituting the above result into \autoref{beam}, we obtain $b=\left(\cos^2 \alpha/4\gamma \right) R_{\ast}/R_{\rm in}$. Expressing $\mu=BR_*^{3}/2$ in terms of the polar surface field strength $B$, it follows from \autoref{R_in} that
\begin{equation}
R_{\rm in}= \left( \frac{\gamma \sqrt{GM_{\ast}}B^{2}R_{\ast}^{4}\delta}{4\pi d^2 \cos^2 \alpha \ F_{\rm X}}  \right)^{2/5}
\label{Rinsx}
\end{equation}
for the inner disk radius and
\begin{equation}
b = \left(\frac{\pi d^2 \cos^7\alpha \ F_{\rm X}}{8\gamma^{7/2} \sqrt{GM_{\ast}} B^2 R_{\ast}^{3/2}\delta} \right)^{2/5}
\label{beaming}
\end{equation}
for the beaming fraction, which can be used in \autoref{dotM} to find
\begin{equation}
\dot{M}_* = \left[ \frac{4\pi^{7/2} d^7 \cos^7 \alpha \ R_{\ast} F_{\rm X}^{7/2}}{\gamma^{7/2} \left(GM_{\ast} \right)^3 B^2\delta} \right]^{2/5}.
\label{dotM2}
\end{equation}
Note from \autoref{dotM2} that the mass accretion rate, $\dot{M}_*$, and the X-ray flux, $F_{\rm X}$, are not linearly proportional as a consequence of beaming. This is because a higher $\dot{M}_*$ results in a smaller inner radius, which increases the polar cap area and hence the beaming fraction.

The fastness parameter defined by \autoref{fstns} can also be expressed in an explicit form with the use of \autoref{Rinsx} for the inner disk radius and $R_{\rm co} = (GM_{\ast}/\Omega_*^2)^{1/3}$ for the corotation radius. We find
\begin{equation}
\omega_*=\left(\frac{\gamma \sqrt{GM_*}B^2 R_*^4\delta}{4\pi d^2\cos^2\alpha F_{\rm X}}\right)^{3/5} \frac{2\pi}{P\sqrt{GM_*}} \label{exfst}
\end{equation}
for the fastness parameter. Here, $P=2\pi/\Omega_*$ is the spin period of the neutron star.

Using \autoref{dotM} and \autoref{beaming}, we write the beaming fraction in terms of the mass accretion rate onto the neutron star and model parameters as
\begin{equation}
b\simeq 10^{-3}\left(\frac{\cos^2 \alpha}{\gamma}\right)\left(\frac{M_{1.4} \dot{M}_{18}^{2}}{R_{10}^{5} B_{13}^{4} \delta_{0.1}^{2}}\right)^{1/7}, \label{exbmn}
\end{equation}
where $M_{1.4}\equiv M_*/1.4\,\Msun$, $R_{10}\equiv R_*/10\,{\rm km}$, $\dot{M}_{18}\equiv \dot{M}_*/10^{18}\,{\rm g\,s^{-1}}$, $B_{13}\equiv B/10^{13}\,{\rm G}$, and $\delta_{0.1}\equiv \delta/0.1$. In \autoref{bmng}, we display the variation of the beaming fraction with $\gamma$ (normalization constant) for the wide ranges of $\dot{M}_*$ and $B$. Given a certain value of $\gamma$, the systems with low $\dot{M}_*$ and high $B$ have lower beaming fractions compared to those with high $\dot{M}_*$ and low $B$.

In the absence of a direct observational clue, about the inclination angle $\alpha$ and the normalization constant $\gamma$, the magnetic field estimation based on these uncertain model parameters might be deceptive. We will express the magnetic field and fastness parameter in terms of beaming fraction without referring to $\alpha$ and $\gamma$ (see \autoref{Bitobw} and Section~\ref{results_beaming}).

Using \autoref{beaming} and \autoref{exfst}, we eliminate $\gamma$ to obtain
\begin{equation}
B=\frac{2^{5/6} \left(GM_*\right)^{1/3} \left(P\omega_*\right)^{7/6} d\sqrt{F_{\rm X}b}}{\pi^{2/3} R_*^{5/2} \sqrt{\delta}} \label{Bitobw}
\end{equation}
for the dipolar magnetic field strength on the surface of the neutron star.

\section{Inference of Magnetic Field From Torque and Luminosity} \label{results_beaming}

In this section, we estimate the magnetic field strengths of PULX using different approaches based on the accretion torque acting on the neutron star and the maximum critical luminosity limit on the observed X-ray luminosity. Throughout our analysis, we usually employ $M_* =1.4\,M_\odot$, $R_* =10\,{\rm km}$, and $I=10^{45}\,{\rm g\,cm^2}$ for the neutron-star mass, radius, and moment of inertia, respectively. As we will see in the case of NGC~1313~X-2 and Swift~J0243.6+6124, however, we will use different values for the neutron-star mass to be able to find at least a marginal solution for NGC~1313~X-2 for which we infer magnetic field from spin-up rate alone and for Swift~J0243.6+6124 for which the magnetic field is\,\,\,\,inferred\,\,\,from\,\,\,spin-up\,\,\,\,rate\,\,\,and\,\,\,critical\,\,\,luminosity.

\begin{deluxetable*}{lccccc}
\tablecolumns{6} \tablewidth{0pt} \tablecaption{Observed periods, period derivatives (spin-up), X-ray fluxes, and distances of PULX.\label{tab:eq}}
\tablehead{ \colhead{Source}\hspace{2.85cm} & \colhead{$P\left(\rm s\right)$\hspace{0.65cm}} & \colhead{$\dot{P}\left(\rm s\ s^{-1}\right)$\hspace{0.65cm}} & \colhead{$F_{\rm X} \left(\rm erg \ s^{-1} cm^{-2}\right)$\hspace{0.65cm}} & \colhead{$d\left(\rm Mpc\right)$\hspace{0.65cm}} & \colhead{References}} \startdata
ULX NGC 5907 & \hspace{0.65cm}1.137\hspace{0.65cm} & \hspace{0.65cm}$-5\times10^{-9}$\hspace{0.65cm} & \hspace{0.65cm}$4.29\times 10^{-12}$\hspace{0.65cm} & \hspace{0.65cm}17.1\hspace{0.65cm} & (1) \\
ULX NGC 7793 P13 & \hspace{0.65cm}0.417\hspace{0.65cm} & \hspace{0.65cm}$-2\times10^{-12}$\hspace{0.65cm} & \hspace{0.65cm}$5.19\times 10^{-12}$\hspace{0.65cm} & \hspace{0.65cm}3.9\hspace{0.65cm} & (2) \\
M82 X-2 (J095551$+$6940.8) & \hspace{0.65cm}1.37\hspace{0.65cm} & \hspace{0.65cm}$-2\times10^{-10}$\hspace{0.65cm} & \hspace{0.65cm}$4.18\times 10^{-12}$\hspace{0.65cm} & \hspace{0.65cm}3.6\hspace{0.65cm} & (3) \\
NGC 300 ULX1 & \hspace{0.65cm}31.6\hspace{0.65cm} & \hspace{0.65cm}$-5.56\times 10^{-7}$\hspace{0.65cm} & \hspace{0.65cm}$1.11\times 10^{-11}$\hspace{0.65cm} & \hspace{0.65cm}1.88\hspace{0.65cm} & (4) \\
M51 ULX-7 & \hspace{0.65cm}2.798\hspace{0.65cm} & \hspace{0.65cm}$-2.4\times 10^{-10}$\hspace{0.65cm} & \hspace{0.65cm}$7.95\times 10^{-13}$\hspace{0.65cm} & \hspace{0.65cm}8.58\hspace{0.65cm} & (5) \\
NGC 1313 X-2 & \hspace{0.65cm}1.458\hspace{0.65cm} & \hspace{0.65cm}$-1.38\times 10^{-8}$\hspace{0.65cm} & \hspace{0.65cm}$9.43\times 10^{-12}$\hspace{0.65cm} & \hspace{0.65cm}4.2\hspace{0.65cm} & (6) \\
Swift J0243.6+6124 & \hspace{0.65cm}9.833\hspace{0.65cm} & \hspace{0.65cm}$-2.3\times 10^{-8}$\hspace{0.65cm} & \hspace{0.65cm}$2.90\times 10^{-7}$\hspace{0.65cm} & \hspace{0.65cm}0.007\hspace{0.65cm} & (7) \\
\enddata
\tablerefs{(1) \citealt{isr+16a}; (2) \citealt{fue+16}; (3) \citealt{bach+14}; (4) \citealt{carp+18};\\(5) \citealt{rod+20}; (6) \citealt{sat+19}; (7) \citealt{wil+18}.}
\end{deluxetable*}

\noindent We will also consider two different values for the neutron-star mass in M51~ULX-7, namely $M_*=1.4M_\odot$ and $M_*=2M_\odot$, to discuss the effect of mass on the model curves. We will briefly mention the effect of the neutron-star radius on the model prediction of magnetic field as well. We use the empirical relation proposed by \citet{lat05} to calculate the neutron-star moment of inertia for different values of mass and radius \citep[see also][]{wor08}.

\subsection{Magnetic Field Inferred From Spin-Up Rate} \label{bfrmsu}

We write the accretion torque acting on the neutron star as $N = n\dot{M}_*\sqrt{GM_{\ast} R_{\rm in}}$, where $n$ is the dimensionless torque. Using the observed spin periods, $P$, and period derivatives, $\dot{P}$, we solve the torque equation $N=I\dot{\Omega}_*$ to estimate the magnetic field of the neutron star of moment of inertia $I$. If the system is away from spin equilibrium, the dimensionless torque can be approximated as $n=n_0$, where $n_0$ is a constant of order unity. Next, we employ \autoref{fstns} to write the torque equation as
\begin{equation}
-2\pi I \dot{P}/P^2 = \omega_{\ast}^{1/3}n_0\sqrt{GM_{\ast} R_{\rm co}}\dot{M}_*.
\label{torque1}
\end{equation}
Solving \autoref{exfst} for $B$, it follows from \autoref{dotM2} that
\begin{equation}
\dot{M}_* =\frac{\pi R_{\ast}^2 F_{\rm X} d^2 \cos^2 \alpha }{\gamma \left(GM_{\ast}\right)^{4/3}}
\left( \frac{2\pi}{\omega_* P} \right)^{2/3},
\label{dotM3}
\end{equation}
which we exploit in \autoref{torque1} to obtain
\begin{equation}
\gamma=\left(\frac{\pi}{4}\right)^{1/3} \frac{n_0 F_{\rm X} P^{5/3} \left(R_* d \cos \alpha \right)^2}{I|\dot{P}|\left(GM_*\right)^{2/3} \omega_*^{1/3}}.
\label{gama}
\end{equation}
Substituting \autoref{gama} into \autoref{exfst} and \autoref{beaming}, we estimate the magnetic field and the fastness parameter in terms of the beaming fraction and the measured quantities $P$, $\dot{P}$, $F_{\rm X}$, and $d$ in \autoref{tab:eq}. We find
\begin{equation}
B=\frac{\omega_*}{R_*^3}\sqrt{\frac{2GM_*I|\dot{P}|}{\pi n_0\delta}} \label{Bsu}
\end{equation}
and
\begin{equation}
\omega_*=\frac{\pi GM_*}{4}\left(\frac{I|\dot{P}|}{n_0 F_{\rm X} d^2 P^{7/3} R_* b}\right)^{3}.\label{fastsu}
\end{equation}
The expressions in \autoref{Bsu} and \autoref{fastsu} can also be obtained substituting \autoref{dotM} and \autoref{Bitobw} into \autoref{torque1}. Here, the mass and radius dependence of the magnetic field and fastness parameter can be roughly estimated as $B\propto M_{*}^5 R_*$ and $\omega_*\propto M_{*}^4 R_{*}^3$ assuming $I\propto M_* R_{*}^2$ to zeroth order. Both quantities, but in particular $B$ is so sensitive to mass that increasing $M_*$ by a factor of two causes $B$ to increase by more than one order of magnitude. The fastness, on the other hand, is sensitive to both mass and radius. For a given mass, $\omega_*$ can decrease by a factor of two if $R_*$ decreases from $\sim13\,{\rm km}$ to $\sim10\,{\rm km}$.

\begin{figure*}
    \centering
            \includegraphics[width=\textwidth]{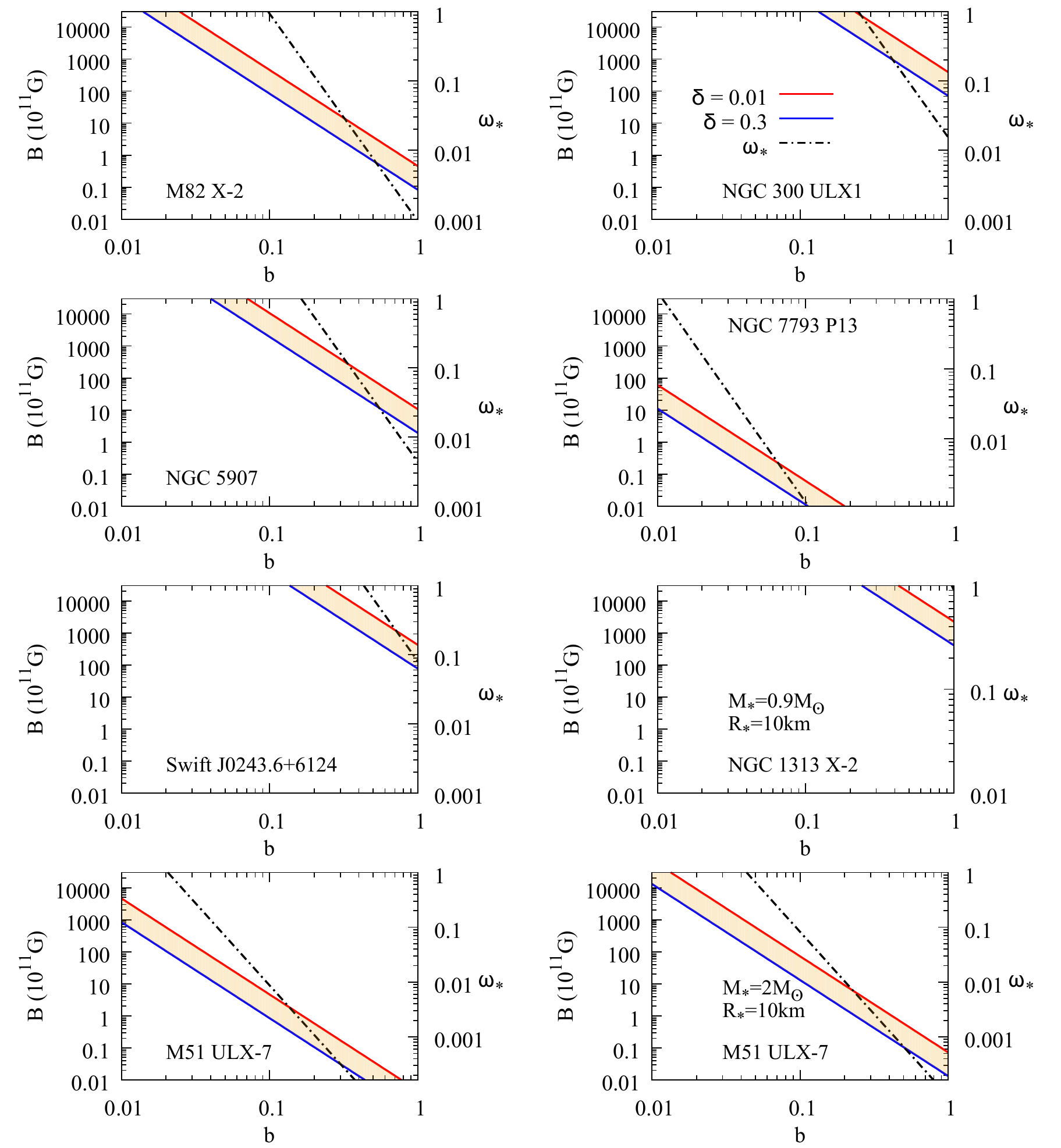}             
    \caption{Estimation of the magnetic field (left vertical axis on each panel) and the fastness parameter (right vertical axis on each panel) in terms of the beaming fraction (horizontal axis on each panel) using the spin-up rates of PULX. The red and blue solid curves corresponding to $\delta =0.01$ and $\delta =0.3$, respectively, determine the upper and lower limits of the allowed range for the magnetic field (shaded region). The run of the fastness parameter is shown by the dot dashed curve. Unless otherwise stated (e.g. NGC~1313~X-2, see Section~\ref{bfrmsu}), we used $M_*=1.4M_\odot$ and $R_*=10\,{\rm km}$ for all sources. The bottom left and right panels for M51~ULX-7 compare two different cases with $M_*=1.4M_\odot$ and $M_*=2M_\odot$, respectively.}
   \label{Bwfromsu}
\end{figure*}

In \autoref{Bwfromsu}, we display the possible range of $B$ (shaded region on each panel) for each source determined by the minimum and maximum values of the boundary-region width. The width of the magnetic boundary region is expected to vary from $\delta \simeq 0.01$ to $\delta \simeq 0.3$ \citep[see, e.g.,][]{erk17}. The weak magnetic field range can be estimated by taking the smallest value of $\omega_*$. The lower limit of $b$, which corresponds to the highest fastness parameter $(\omega_* =1)$, yields the strong magnetic field range.

For each source, the minimum value of the fastness parameter is $\omega_{*,{\rm min}}=(R_*/R_{\rm co})^{3/2}$ as $R_{\rm in} >R_*$ (see \autoref{fstns}), corresponding to $b\lesssim 1$ as the disk reaches the neutron star surface. For some sources, \autoref{fastsu}, with nominal values of the parameters $n_0, I, M_*,$ and $R_*$

\begin{deluxetable*}{lcc}
\tablecolumns{3} \tablewidth{0pt} \tablecaption{Weak (min) and strong (max) magnetic field ranges inferred from spin-up rates. \label{tab:spinup}}
\tablehead{ \colhead{Source}\hspace{4cm} & \colhead{$B_{\rm min}\left(\times10^{11}\,{\rm G}\right)$\hspace{4cm}} & \colhead{$B_{\rm max}\left(\times10^{13}\,{\rm G}\right)$}} \startdata
ULX NGC 5907 & \hspace{4cm}2.0--10\hspace{4cm} & 4.0--25 \\
ULX NGC 7793 P13 & \hspace{4cm}0.012--0.070\hspace{4cm} & 0.09--0.5 \\
M82 X-2 (J095551$+$6940.8) & \hspace{4cm}0.08--0.5\hspace{4cm} & 0.80--5.0 \\
NGC 300 ULX1 & \hspace{4cm}70--400\hspace{4cm} & 45--240 \\
M51 ULX-7 & \hspace{4cm}0.018--0.090\hspace{4cm} & 1.0--6.0 \\
NGC 1313 X-2 & \hspace{4cm}400--2300\hspace{4cm} & 4.6--25 \\ 
Swift J0243.6+6124 & \hspace{4cm}80--420\hspace{4cm} & 10--50 \\  
\enddata
\tablecomments{Here, the magnetic field ranges of NGC~1313~X-2 are marginally obtained at $b\simeq1$ for $M_*=0.9\,M_{\odot}$ and $R_*=10\,{\rm km}$\\(see Section~\ref{bfrmsu}). For all other sources, we use $M_*=1.4M_\odot$ and $R_*=10\,{\rm km}$.}
\end{deluxetable*}

\noindent and the mean observational value of the distance $d$ yields an upper limit for the beaming factor, $b_{\rm max} <1$ corresponding to $\omega_{*,{\rm min}}$. For other sources we find $\omega_*(b=1)>\omega_{*,{\rm min}}$ from \autoref{fastsu}. With actual values of the parameters these sources would also give an upper bound $b_{\rm max}\leq 1$. The strong magnetic fields obtain at the highest value of the fastness parameter, $\omega_{*} =1$, and the corresponding beaming factor, $b_{\rm min} < 1$, while the weak magnetic field range obtains at beaming factor, $b\lesssim 1$, with $\omega_{*,{\rm min}} < 1$, the case of isotropic radiation when the accretion disk reaches down to the neutron star surface. In the case of M82~X-2, the lower limit of the beaming fraction can be identified at $\omega_* =1$ as $b\simeq 0.1$, which corresponds to the strong magnetic field range of $8\times 10^{12}\,{\rm G}$ to $5\times 10^{13}\,{\rm G}$. The upper limit of the beaming fraction $(b=1)$ determines the weak magnetic field range of $8\times 10^9\,{\rm G}$ to $5\times 10^{10}\,{\rm G}$ (\autoref{Bwfromsu}). ULX~NGC~5907 and NGC~300~ULX1 also allow for solutions from $\omega_*=1$ to $b=1$ allowing magnetic fields from the $10^{14}-10^{15}\,{\rm G}$ strong range down to the $10^{11}-10^{13}\,{\rm G}$ weak range. The strong field range in Swift~J0243.6+6124 corresponds to a lower limit $b\simeq 0.44$ for the beaming fraction. The fastest rotating PULX (ULX~NGC~7793~P13) is one of two sources in which the beaming fraction has a maximum value $b\simeq 0.1$ in accordance with $\omega_* >(R_*/R_{\rm co})^{3/2}$, that is, $\omega_* \gtrsim 0.001$ for the observed period ($P\simeq 0.4\,{\rm s}$, see \autoref{tab:eq}) of this source. The upper limit of the beaming fraction $(b\simeq 0.1)$ corresponds to the weak magnetic field range of $(1.2-7)\times 10^9\,{\rm G}$. For $\omega_* <1$, we read the lower limit of the beaming fraction as $b\gtrsim 0.01$, which determines the strong magnetic field range of $9\times 10^{11}-5\times 10^{12}\,{\rm G}$, as shown in \autoref{Bwfromsu}. We estimate the upper limit of the beaming fraction in M51~ULX-7 as $b\simeq 0.37$, which determines the weak magnetic field range for $M_*=1.4M_\odot$ and $R_*=10\,{\rm km}$.  Note in M51~ULX-7 that the fastness parameter and magnetic field values for $M_*=1.4M_\odot$ (bottom left panel of \autoref{Bwfromsu}) are at least one order of magnitude smaller at a given $b$ than the corresponding values for $M_*=2M_\odot$ (bottom right panel of \autoref{Bwfromsu}). As seen in the case of M51~ULX-7, the smaller the neutron-star mass is, the lower the fastness parameter can be. In the absence of any solution with $\omega_*<1$ for NGC~1313~X-2 when $M_*=1.4M_\odot$ and $R_*=10\,{\rm km}$, we obtain a marginal solution at $b\simeq 1$ and $\omega_*\simeq 1$ keeping both the mass and radius of the neutron star as small as $M_*=0.9M_\odot$ and $R_*=10\,{\rm km}$, respectively. We summarize our results for the magnetic field ranges inferred from spin-up rates of PULX in \autoref{tab:spinup}.

\subsection{Magnetic Field Inferred From Spin Equilibrium} \label{Bfrmeq}

The short spin-up time scales $P/\dot{P}\sim 100~{\rm yrs}$ of the PULX \citep{bach+14,isr+16a,isr+16b,fue+16} imply that the neutron stars in these systems must be close to spin equilibrium with their disks. The assumption of spin equilibrium, of course, would not be justified if the disk is in outburst corresponding to an enhanced accretion rate at the time of observation as is possibly the case with the PULX in M82 \citep{tsy+16}.

The system is assumed to reach spin equilibrium at a critical fastness parameter, $\omega_{\rm c} \lesssim 1$, as all torque models near spin equilibrium behave as $n\propto1-\omega_{\ast} / \omega_{\rm c}$. At the spin equilibrium, $\omega_*=\omega_{\rm c}$ and
the inner radius of the disk is related to the corotation radius through
$R_{\rm in}=\omega_{\rm c}^{2/3}R_{\rm co}$. Accordingly, $B$ can be deduced as a function of $b$ from \autoref{Bitobw} for $\omega_*=\omega_{\rm c}$. The mass and radius dependence of the magnetic field can be read from \autoref{Bitobw} as $B\propto M_{*}^{1/3} /R_{*}^{5/2}$. For a given mass, $B$ decreases by a factor of two if $R_*$ increases from \,$\sim10\,\,{\rm km}$ \,to \,$\sim13\,\,{\rm km}$. \,In \,general, \,we \,infer \,lower

\begin{figure*}
    \centering
            \includegraphics[width=\textwidth]{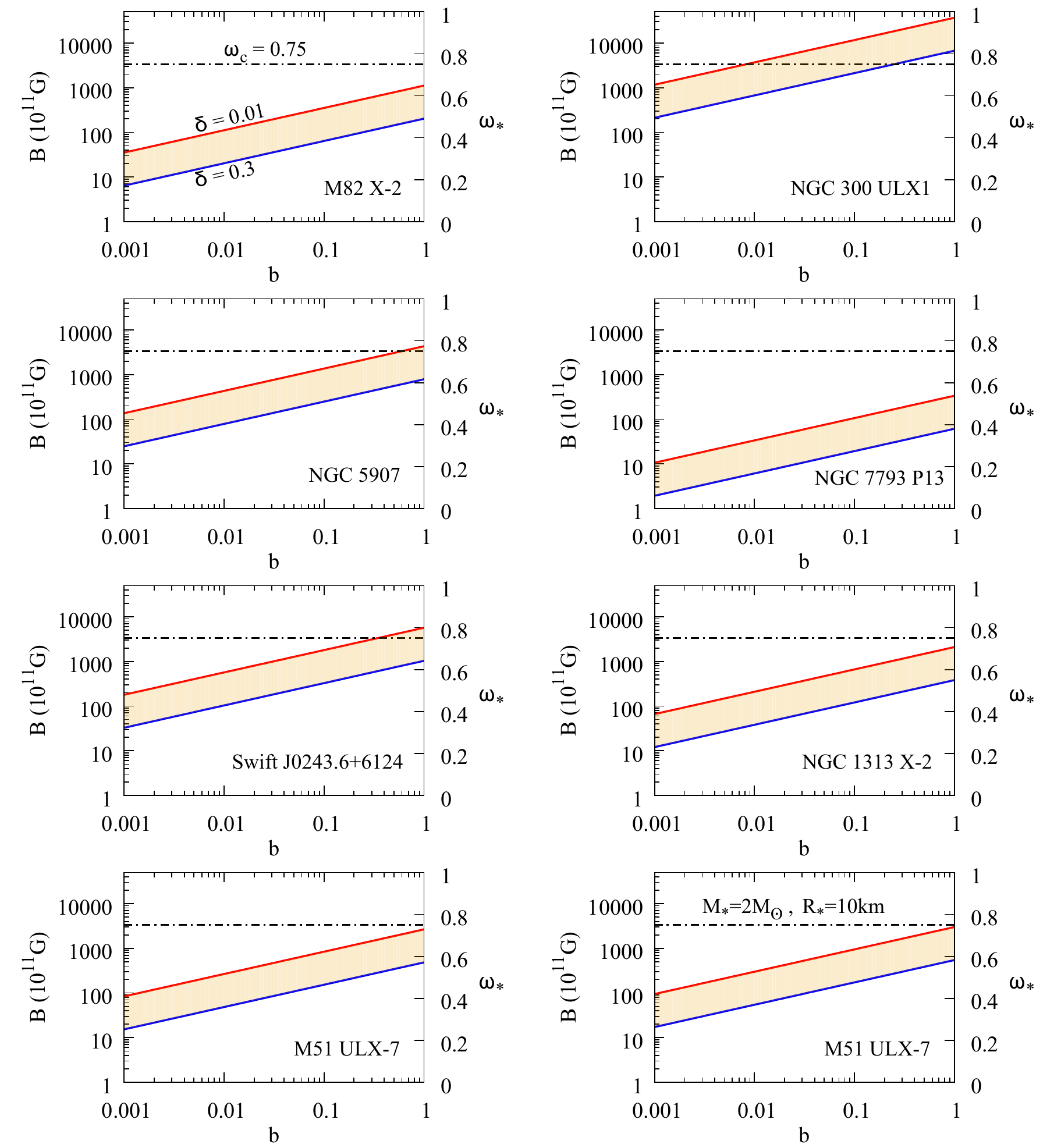}             
    \caption{Estimation of the magnetic field (left vertical axis on each panel) and the fastness parameter (right vertical axis on each panel) in terms of the beaming fraction (horizontal axis on each panel), provided PULX are close to spin equilibrium. The fastness parameter (dot dashed line) is assumed to have its critical value, i.e., $\omega_*=\omega _{\rm c}=0.75$. The red and blue solid curves corresponding to $\delta =0.01$ and $\delta =0.3$, respectively, determine the upper and lower limits of the allowed range for the magnetic field (shaded region). We use $M_*=1.4M_\odot$ and $R_*=10\,{\rm km}$ for all sources. The bottom left and right panels for M51~ULX-7 compare two different cases with $M_*=1.4M_\odot$ and $M_*=2M_\odot$, respectively (see Section~\ref{Bfrmeq} for the effect of larger neutron-star radius on the magnetic field).}
   \label{Bwfromeq}
\end{figure*}

\begin{deluxetable*}{lcc}
\tablecolumns{3} \tablewidth{0pt} \tablecaption{Weak (min) and strong (max) magnetic field ranges inferred from spin equilibrium. \label{tab:spineq}}
\tablehead{ \colhead{Source}\hspace{4cm} & \colhead{$B_{\rm min}\left(\times 10^{13}\,{\rm G}\right)$\hspace{4cm}} & \colhead{$B_{\rm max}\left(\times 10^{14}\,{\rm G}\right)$}} \startdata
ULX NGC 5907 & \hspace{4cm}0.24--1.4\hspace{4cm} & 0.80--4.5 \\
ULX NGC 7793 P13 & \hspace{4cm}0.077--0.42\hspace{4cm} & 0.061--0.33 \\
M82 X-2 (J095551$+$6940.8) & \hspace{4cm}0.30--1.7\hspace{4cm} & 0.20--1.0 \\
NGC 300 ULX1 & \hspace{4cm}12--67\hspace{4cm} & 7.0--35 \\
M51 ULX-7 & \hspace{4cm}0.70--3.9\hspace{4cm} & 0.50--2.6 \\
NGC 1313 X-2 & \hspace{4cm}0.35--1.8\hspace{4cm} & 0.40--2.1 \\ 
Swift J0243.6+6124 & \hspace{4cm}3.2--17\hspace{4cm} & 1.0--5.7 \\ 
\enddata
\tablecomments{For all sources, we use $M_*=1.4M_\odot$ and $R_*=10\,{\rm km}$ (see Section~\ref{Bfrmeq} for the effect of larger $R_*$ on $B$).}
\end{deluxetable*}

\noindent values of $B$ from spin equilibrium for neutron stars of relatively small masses and large radii.

In \autoref{Bwfromeq}, we estimate $B$, keeping the fastness parameter constant at the critical value, $\omega _{\rm c} =0.75$ \citep{tur+17}, which would be appropriate for PULX that are sufficiently close to spin equilibrium.

Unlike the inference of magnetic field from spin-up rates, the strong magnetic field range is determined by the upper limit of the beaming fraction $(b=1)$. We consider the super-Eddington rate of mass transfer to guess the lower limit of $b$ and hence the weak magnetic field range. Using \autoref{mdot}, we write
\begin{equation}
\dot{M}_0=\left(\frac{R_{\rm sp}}{\omega_{\rm c}^{2/3}R_{\rm co}}\right)\dot{M}_* >\frac{L_{\rm E}}{\epsilon c^2}, \label{supedacc}
\end{equation}
where $R_{\rm sp}=27\epsilon GM_* \dot{M}_0 /2L_{\rm E}$. Substituting \autoref{dotM} for $\dot{M}_*\equiv \dot{M}\left(R_{\rm in}\right)$, it follows from \autoref{supedacc} that
\begin{equation}
b\gtrsim 0.07\,\frac{P^{2/3}M_{1.4}^{1/3}R_{10}}{\dot{M}_{20}F_{11} d_1^2} \label{lowb}
\end{equation}
provided $\omega_{\rm c} =0.75$ \citep{tur+17}. Here, $P$ is the spin period in seconds, $F_{11}$ is the X-ray flux in units of $10^{-11}\,{\rm erg\,s^{-1}\,cm^{-2}}$, $d_1$ is the distance in Mpc, $\dot{M}_{20}$ is $\dot{M}_0$ in units of $10^{20}\,{\rm g\,s^{-1}}$, $M_{1.4}$ is $M_*$ in units of $1.4\,M_\odot$, and $R_{10}$ is $R_*$ in units of $10\,{\rm km}$. In order to be self-consistent with \autoref{mdot} and \autoref{lowb}, $R_{\rm in}<R_{\rm sp}$. The minimum rate of mass transfer that guarantees $R_{\rm in}<R_{\rm sp}$ can be guessed through $R_{\rm co}\lesssim R_{\rm sp}$ as $R_{\rm in}=\omega_{\rm c}^{2/3}R_{\rm co}<R_{\rm co}$ for $\omega_{\rm c}=0.75$. We expect from $R_{\rm co}\approx R_{\rm sp}$ that $P^{2/3}/\dot{M}_{20}\approx 1.8\,M_{1.4}^{2/3} /R_{10}$ and therefore
\begin{equation}
b\gtrsim 0.13\,F_{11}^{-1} d_1^{-2} M_{1.4}. \label{lwlimb}
\end{equation}
This lower limit also holds for $R_{\rm in}>R_{\rm sp}$ with $\dot{M}_0=\dot{M}_{\ast}$. We employ the observed values of $F_{\rm X}$ and $d$ in \autoref{tab:eq} and estimate this minimum value of $b$ for each PULX. The weak and strong field ranges that correspond to the minimum $b$ and with $b=1$, respectively, can be read from \autoref{Bwfromeq}. In \autoref{tab:spineq}, we present our results for the magnetic field ranges inferred from spin equilibrium.

\begin{figure*}
    \centering
            \includegraphics[width=\textwidth]{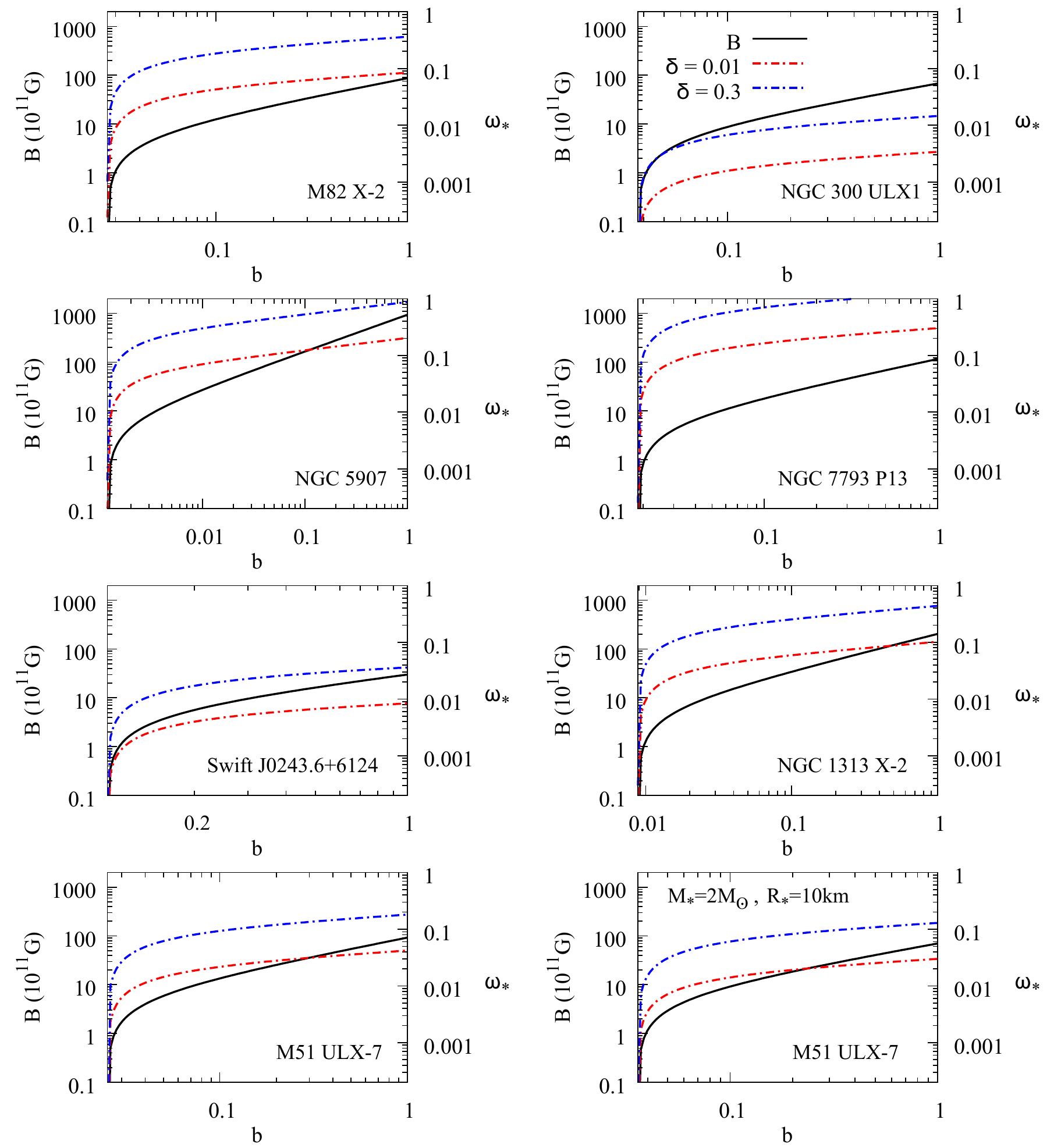}             
    \caption{Estimation of the magnetic field (left vertical axis on each panel) and the fastness parameter (right vertical axis on each panel) in terms of the beaming fraction (horizontal axis on each panel), provided PULX have $L_{\rm X}=L_{\rm c}$. The magnetic field is shown by the black solid curve. The run of the fastness parameter is displayed by the red and blue dot dashed curves corresponding to $\delta =0.01$ and $\delta =0.3$, respectively. We use $M_*=1.4M_\odot$ and $R_*=10\,{\rm km}$ for all sources. The bottom left and right panels for M51~ULX-7 compare two different cases with $M_*=1.4M_\odot$ and $M_*=2M_\odot$, respectively (see Section~\ref{critB}).}
   \label{Bwfromlum}
\end{figure*}

\subsection{Magnetic Field Inferred From Critical Luminosity} \label{critB}

A strong magnetic field can increase the critical luminosity of an accreting pulsar by decreasing the cross-section for scattering of photons from electrons \citep{can71,pac92,mus+16}. The critical luminosity, $L_{\rm c}$, depends on the magnetic field as $311\left(B/B_{\rm c}\right)^{4/3}L_{\rm E}$ if $L_{\rm c} \gg L_{\rm E}$ \citep{pac92,ton15a}, where $B_{\rm c}\equiv m_{\rm e}^{2}c^{3}/\hbar e=4.4\times 10^{13}~{\rm G}$ is the quantum critical magnetic field. For surface field strengths above the quantum critical value, the critical luminosity exceeds the Eddington luminosity by more than two orders of magnitude. For surface field strengths well below the quantum critical value, $L_{\rm c}\simeq L_{\rm E}$. Assuming that the X-ray luminosity is limited by the critical luminosity, that is,

\begin{equation}
L_{\rm X}\leq L_{\rm c} \simeq \left[1+311\left(\frac{B}{B_{\rm c}}\right)^{4/3} \right]L_{\rm E}, \label{lumcr}
\end{equation}
where $L_{\rm X}=4\pi bd^2 F_{\rm X}$, we estimate the surface magnetic field strength for each PULX in terms of the beaming fraction, X-ray flux, and distance to the source as
\begin{equation}
B\gtrsim \frac{B_{\rm c}}{\left(311\right)^{3/4}}\left(\frac{4\pi d^2 F_{\rm X} b}{L_{\rm E}}-1\right)^{3/4}, \label{Bcrlum}
\end{equation}
which in turn can be used in \autoref{Bitobw} to solve for the fastness parameter. Note that \autoref{Bcrlum} yields the minimum magnetic field strength for $L_{\rm X}=L_{\rm c}$.

In \autoref{Bwfromlum}, we display the minimum $B$ when $L_{\rm X}=L_{\rm c}$ and the corresponding fastness parameter for a wide range of beaming. Unlike the inference of $B$ from spin-up \,rates \,and \,spin \,equilibrium, \,the \,magnetic \,field \,has

\begin{deluxetable*}{lcc}
\tablecolumns{3} \tablewidth{0pt} \tablecaption{Lowest and highest values for the minimum magnetic field ranges inferred from critical luminosity.\label{tab:crlum}}
\tablehead{ \colhead{Source}\hspace{4cm} & \colhead{$B_{\rm low}\left(\times 10^{9}\,{\rm G}\right)$\hspace{4cm}} & \colhead{$B_{\rm high}\left(\times 10^{13}\,{\rm G}\right)$}} \startdata
ULX NGC 5907 & \hspace{4cm}0.40--2.3\hspace{4cm} & 9.0 \\
ULX NGC 7793 P13 & \hspace{4cm}0.40--2.2\hspace{4cm} & 0.45--1.2 \\
M82 X-2 (J095551$+$6940.8) & \hspace{4cm}0.40--2.2\hspace{4cm} & 0.85 \\
NGC 300 ULX1 & \hspace{4cm}0.40--2.2\hspace{4cm} & 0.67 \\
M51 ULX-7 & \hspace{4cm}0.40--2.3\hspace{4cm} & 0.92 \\
NGC 1313 X-2 & \hspace{4cm}0.40--2.2\hspace{4cm} & 2.0 \\ 
Swift J0243.6+6124 & \hspace{4cm}0.40--2.2\hspace{4cm} & 0.30 \\ 
\enddata
\tablecomments{The lowest range corresponds to the range of boundary region thickness $\delta$ values at $\omega_{*,{\rm min}}$. The highest value is the\\value at $b=1$. The case of ULX~NGC~7793~P13 is discussed in the text. For all sources, we use $M_*=1.4M_\odot$ and\\ $R_*=10\,{\rm km}$ (see Section~\ref{critB} for the effect of higher $M_*$ on $B$).}
\end{deluxetable*}

\noindent no explicit dependence on $\delta$ (\autoref{Bcrlum}), whereas $\omega_*$ depends on $\delta$ through \autoref{Bitobw}. The upper limit for the beaming fraction $(b=1)$ determines the highest value of the minimum magnetic field for $\omega_*<1$. For each PULX, except ULX~NGC~7793~P13, $\omega_*<1$ for $b=1$ across the $\delta$ range $(0.01<\delta<0.3)$, therefore $b=1$ yields the highest value of the minimum magnetic field. For ULX~NGC~7793~P13, however, $\omega_*<1$ for the upper limit of $\delta$ $(\delta=0.3)$ only if $b<0.3$ (\autoref{Bwfromlum}). The upper limit for the beaming fraction is then given by a set of values $(0.3<b<1)$ in the 0.01--0.3 range of $\delta$, which establishes the range for the highest value of the minimum $B$ for this source. In all sources, the range for the lowest value of the minimum magnetic field is determined by a set of $b$ values (lower limit for the beaming fraction) for which $\omega_* >(R_*/R_{\rm co})^{3/2}$. Our results for the magnetic field ranges inferred from critical luminosity $(L_{\rm X}=L_{\rm c})$ are summarized in \autoref{tab:crlum}. The field ranges for all sources in \autoref{tab:crlum} are obtained for $M_* =1.4\,M_\odot$ and $R_* =10\,{\rm km}$. For a $2M_\odot$ neutron star, as in the second case of M51~ULX-7 we depict at the bottom right panel of \autoref{Bwfromlum}, slightly lower field values and smaller fastness parameters are inferred compared to the bottom left panel of the same figure, because the minimum $B$ decreases as the Eddington luminosity increases with mass (see~\autoref{Bcrlum}).

\subsection{Magnetic Field Inferred From Spin-Up Rate and Critical Luminosity} \label{sucr}

\begin{figure*}
    \centering
            \includegraphics[width=\textwidth]{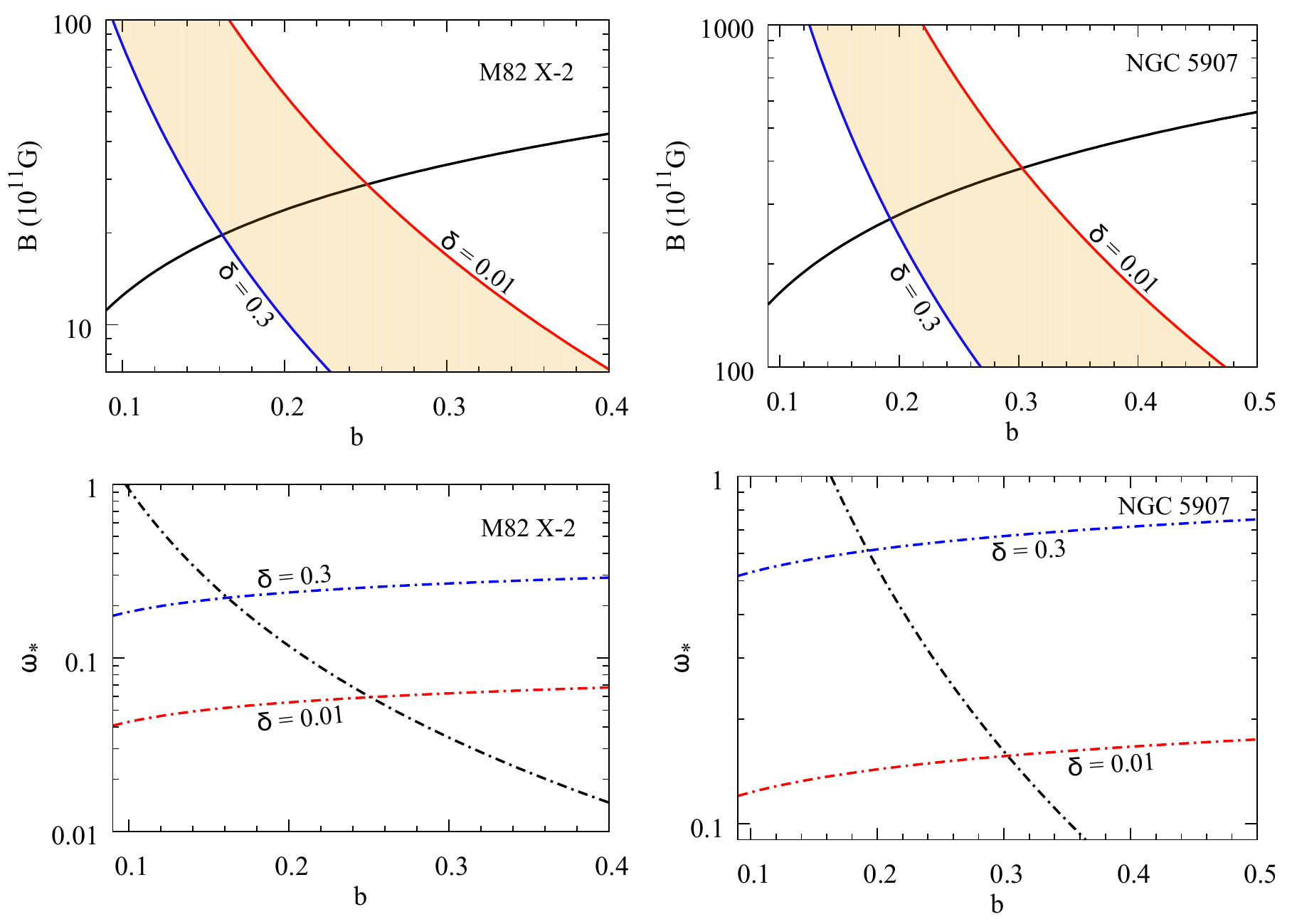}             
    \caption{Estimation of the magnetic field and the fastness parameter in terms of the beaming fraction using the spin-up rates of the PULX M82~X-2 (left panels) and ULX~NGC~5907 (right panels), provided $L_{\rm X}=L_{\rm c}$ for each source. The common solutions to $B$ (vertical axes on the upper panels) and $b$ (horizontal axis on each panel) are given by the intersection of the black solid curve ($B$ inferred from the critical luminosity assumption) and the shaded region between the red and blue solid curves ($B$ inferred from the spin-up rate). Similarly, the common solution to $\omega_*$ (vertical axes on the lower panels) is determined by the intersection of the black dot dashed curve ($\omega_*$ inferred from the spin-up rate) and the region between the red and blue dot dashed curves ($\omega_*$ inferred from the critical luminosity assumption). We use $M_*=1.4M_\odot$ and $R_*=10\,{\rm km}$ for each source.}
   \label{Bwsulum1}
\end{figure*}

\begin{figure*}
    \centering
            \includegraphics[width=\textwidth]{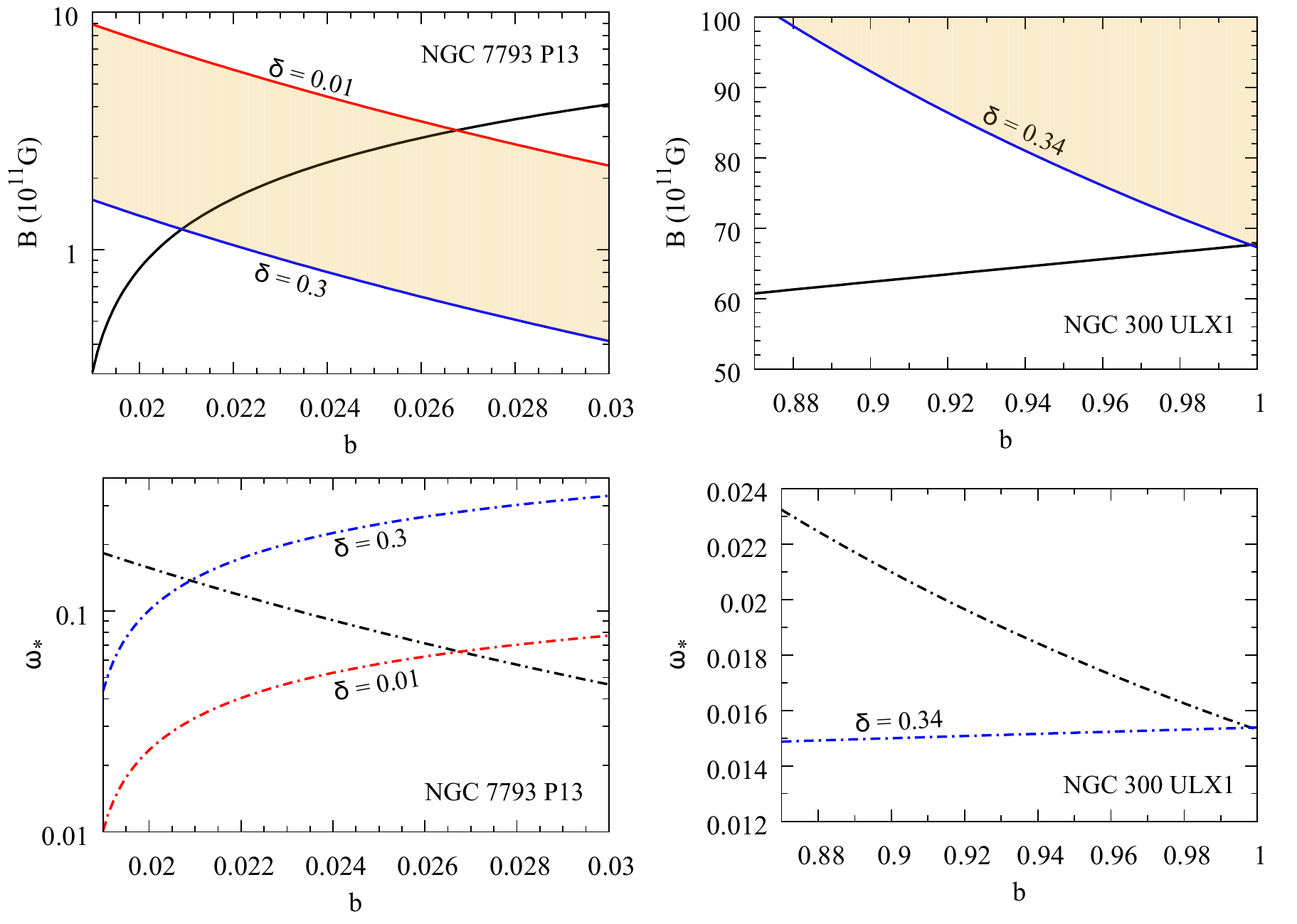}             
    \caption{Same as \autoref{Bwsulum1}, but for ULX~NGC~7793~P13 (left panels) and NGC~300~ULX1 (right panels). We use $M_*=1.4M_\odot$ and $R_*=10\,{\rm km}$ for each source.}
   \label{Bwsulum2}
\end{figure*}

We search for the common solution to the magnetic field inferred from both the spin-up rates and critical luminosity assumption $(L_{\rm X}=L_{\rm c})$. As shown in \autoref{Bwsulum1}, \autoref{Bwsulum2}, and \autoref{Bwsulum3}, we estimate the ranges of magnetic field and fastness parameter through the intersection of model curves we obtain using the observed spin-up rates and the critical luminosity assumption for the X-ray emission. As expected, both $B$ and $\omega_*$ estimates yield the same range for the beaming fraction in each source. In all sources except NGC~300~ULX1, Swift~J0243.6+6124, and NGC~1313~X-2, the model curves based on the spin-up rates and critical luminosities intersect over a range of $b$ to determine the common solution to $B$ and $\omega_*$ between $\delta=0.01$ and $\delta=0.3$ (\autoref{Bwsulum1} and \autoref{Bwsulum2}). In NGC~300~ULX1, the solutions to the magnetic field and the fastness parameter can be marginally found around $b=1$ for $\delta=0.34$, slightly exceeding the upper limit of the boundary region width $(\delta=0.3)$ we employ in the present work. In Swift~J0243.6+6124, the model curves can only intersect for sufficiently small mass and radius of the neutron star. As seen from \autoref{Bwsulum3}, we start finding common solutions to the magnetic field and fastness parameter of Swift~J0243.6+6124 for $M_*\lesssim 1.2M_\odot$ and $R_*=10\,{\rm km}$. In the case of NGC~1313~X-2, however, we found no common solution even for neutron-star masses as small as $0.9M_\odot$. We summarize our results in \autoref{tab:sucrlum}.

\subsection{Magnetic Field Inferred From Spin Equilibrium and Critical Luminosity} \label{eqcr}

As seen from \autoref{Bwfromlum}, the fastness parameter for the PULX M82~X-2, NGC~300~ULX1, M51~ULX-7, NGC~1313~X-2, and Swift~J0243.6+6124 cannot attain values as high as $\omega_{\rm c}=0.75$ if these sources emit X-rays with the maximum critical luminosity. For the PULX ULX~NGC~5907 and ULX~NGC~7793~P13, however, the range of the fastness parameter for $0.01<\delta<0.3$ admits the critical value, $\omega_{\rm c}=0.75$ (\autoref{Bwfromlum}), allowing the possibility that these two sources are close to spin equilibrium while having critical luminosity in X-rays $(L_{\rm X}=L_{\rm c})$. The common solution to $B$ and $\omega_*$ inferred from the intersection of the model curves based on spin equilibrium and critical luminosity assumption can be estimated as shown in \autoref{Bweqlum}. The results we obtain for the beaming fraction and magnetic field strength when $\omega_*=\omega_{\rm c}=0.75$ are summarized in \autoref{tab:eqcrlum}. We could not find any common solutions in the case of other PULX such as M82~X-2, NGC~300~ULX1, M51~ULX-7, NGC~1313~X-2, and Swift~J0243.6+6124 even for the extremely small value of the neutron-star mass, i.e., for $M_*=0.9M_\odot$ while keeping $R_*=10\,{\rm km}$. It is, however, likely that M82~X-2 and NGC~1313~X-2 can still be close to spin equilibrium and emit X-rays at the critical luminosity level for sufficiently small and large values of the neutron-star mass and radius, respectively (e.g., $M_*\lesssim1M_\odot$ and $R_*\gtrsim13\,{\rm km}$). Unlike the case of field inference from spin-up rates and critical luminosity, where both the neutron-star mass and radius must be sufficiently small to find a common solution, the relatively large radii and small masses for the neutron stars in PULX are necessary for both the spin equilibrium and critical luminosity conditions to be satisfied (see also Section~\ref{Bfrmeq}).

\subsection{Magnetic Field Inferred From Subcritical Luminosity}

The lower limit for the magnetic field strength is determined by the critical luminosity (\autoref{Bcrlum}). The ranges of $B$ in \autoref{tab:sucrlum} and \autoref{tab:eqcrlum} thus correspond to the minimum values of the magnetic field inferred at critical luminosity from spin-up rates and spin equilibrium, respectively. If $L_{\rm X}<L_{\rm c}$, which is likely for the PULX, the sources would have stronger fields.

In all PULX with subcritical luminosity $(L_{\rm X}<L_{\rm c})$, the allowed region for the magnetic field and the corresponding ranges for the beaming fraction and fastness parameter are determined by the part of the shaded area (\autoref{Bwfromsu} and \autoref{Bwfromeq}) that lies above the curve of the minimum magnetic field resulting from the critical luminosity assumption (\autoref{Bwfromlum}). We display the allowed regions for the magnetic fields of PULX with subcritical luminosities inferred from spin-up rates and spin equilibrium in \autoref{Bsusub} and \autoref{Beqsub}, respectively.

\subsubsection{Field Inference From Spin-Up Rate at Subcritical Luminosity}

\begin{figure*}
    \centering
            \includegraphics[width=\textwidth]{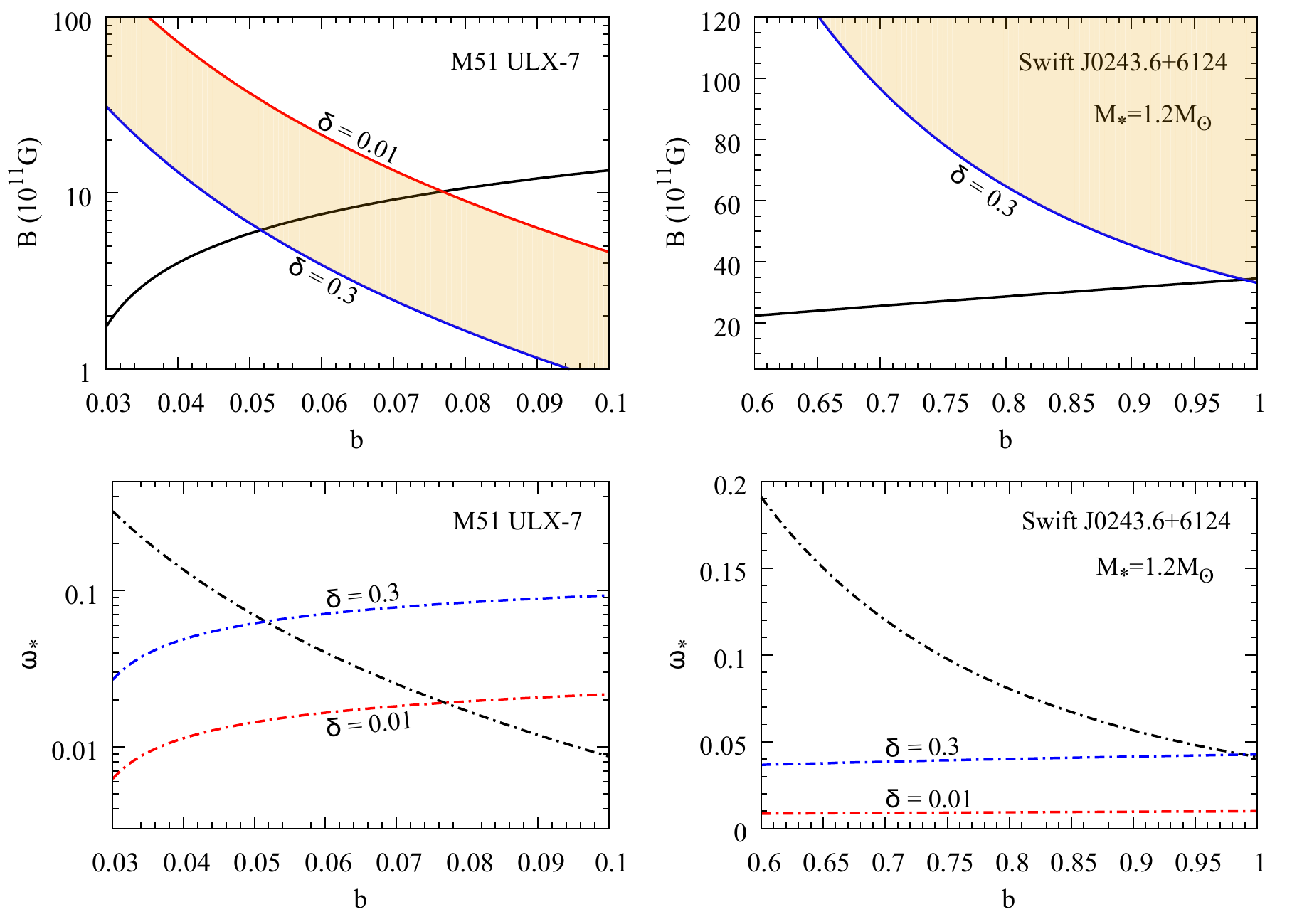}             
    \caption{Same as \autoref{Bwsulum1}, but for M51~ULX-7 (left panels) and Swift~J0243.6+6124 (right panels). We took $R_*=10\,{\rm km}$ for both sources; $M_*=1.4M_\odot$ for M51~ULX-7 and $M_*=1.2M_\odot$ for Swift~J0243.6+6124 (see Section~\ref{sucr}).}
   \label{Bwsulum3}
\end{figure*}

As seen from \autoref{Bwsulum1}, \autoref{Bwsulum2}, and \autoref{Bwsulum3}, the upper limit for the beaming fraction of the sources with subcritical luminosity can be deduced from the intersection of the model curves for $B$ inferred from critical luminosity (curve of the minimum $B$) and spin-up rates for $\delta=0.01$. The intersection of the model curve based on the spin-up rate for $\delta=0.3$ with the minimum $B$ curve yields the lower limit of the magnetic field. The upper limit of $b$ corresponds to the lower limit of the fastness parameter (see \autoref{tab:sucrlum}). The lower limit of $b$, which determines the upper limit of the magnetic field, is found using the upper limit of the fastness parameter $(\omega_*=1)$ (see Section~ \ref{bfrmsu}). In \autoref{Bsusub}, we depict the allowed region for the magnetic field bounded by the minimum $B$ curve, the lower limit of the beaming fraction, and the $B$ curves based on spin-up rates. In \autoref{tab:susclum}, we present our results for the PULX with subcritical luminosities using their spin-up rates.

\subsubsection{Field Inference From Spin Equilibrium at Subcritical Luminosity}

\begin{figure*}
    \centering
            \includegraphics[width=\textwidth]{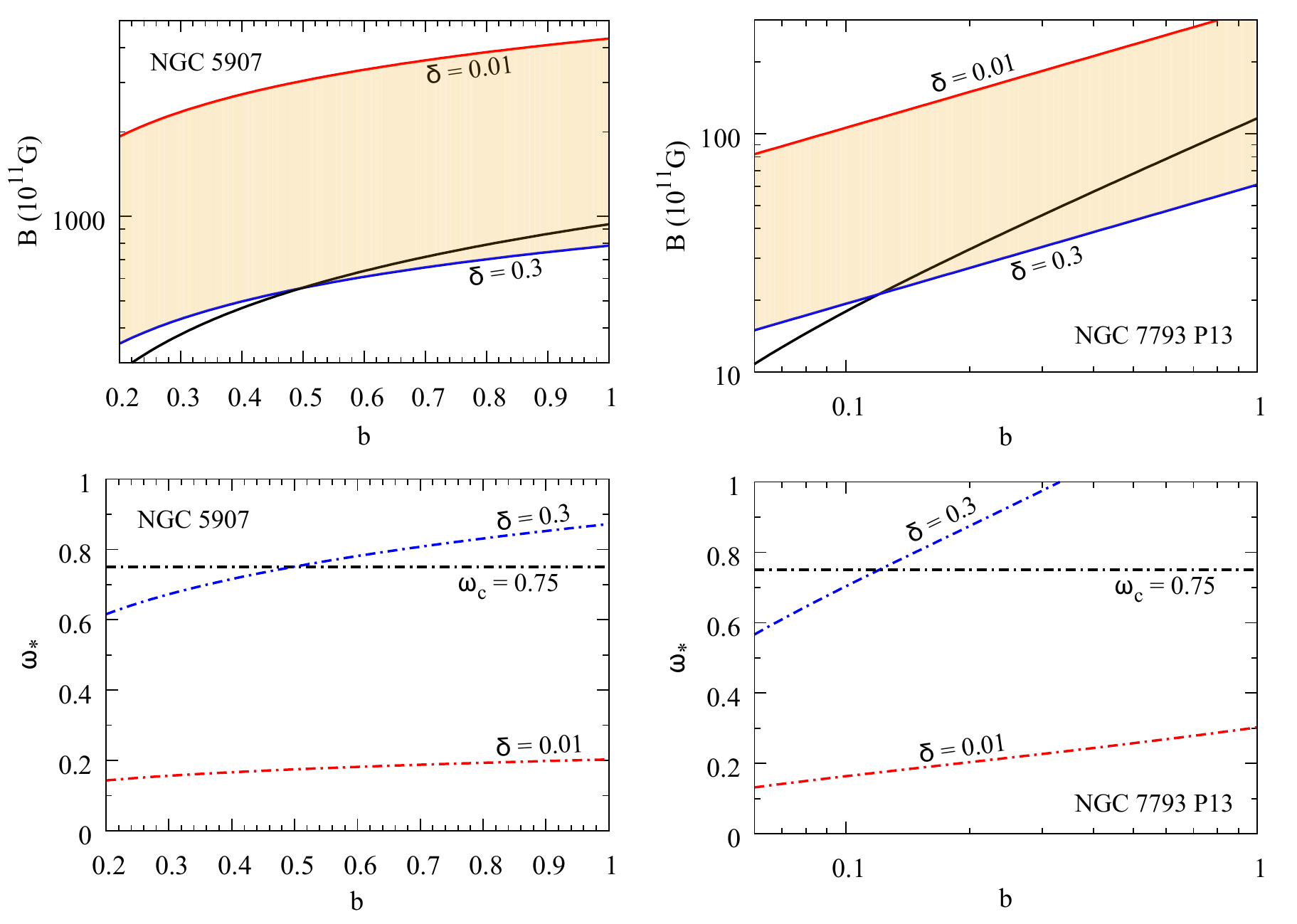}             
    \caption{Estimation of the magnetic field and the fastness parameter in terms of the beaming fraction provided the PULX ULX~NGC~5907 (left panels) and ULX~NGC~7793~P13 (right panels) are close to spin equilibrium and $L_{\rm X}=L_{\rm c}$ for each source. The common solutions to $B$ (vertical axes on the upper panels) and $b$ (horizontal axis on each panel) are given by the section of the black solid curve ($B$ inferred from the critical luminosity assumption) within the shaded region between the red and blue solid curves ($B$ inferred from the spin-equilibrium condition). Similarly, the common solution to $\omega_*$ (vertical axes on the lower panels) is given by the section of the black dot dashed curve ($\omega_{\rm c}=0.75$ is the critical value for the fastness parameter) within the region between the red and blue dot dashed curves for $\omega_*$ inferred from the critical luminosity assumption. We use $M_*=1.4M_\odot$ and $R_*=10\,{\rm km}$ for each source.}
   \label{Bweqlum}
\end{figure*}

Except for the PULX ULX~NGC~5907 and ULX~NGC~7793~P13, one sees from \autoref{Bwfromeq} and \autoref{Bwfromlum} that the minimum $B$ curve determined by the critical luminosity assumption lies well below the shaded area bounded by the model curves of $B$ based on the spin equilibrium condition. For the PULX ULX~NGC~5907 and ULX~NGC~7793~P13, these curves intersect over a certain range of $b$ (see \autoref{Bweqlum}), both the upper and lower limits of the magnetic field range remain within the shaded area boundary above the minimum $B$ curve. The upper and lower limits of the beaming fraction determine the upper and lower limits of the magnetic field, respectively. In \autoref{Beqsub}, we display the allowed region for the magnetic field bounded by the minimum $B$ curve, the lower limit of the beaming fraction, and the $B$ curves for spin equilibrium. We summarize our results for the PULX with subcritical luminosities in \autoref{tab:eqsclum}, assuming these sources are sufficiently close to their spin equilibrium.

\section{Summary and Discussion}
\label{discuss}

The central engines of most if not all ULX are likely to be neutron stars rather than black holes of stellar or intermediate mass \citep{wik+19}. The argument in favor of neutron stars being hosted by a considerable fraction of ULX is supported by the recent discovery of pulsating objects, namely PULX in the zoo of ULX \citep{bach+14}.

\subsection{Case Study and Results}

In this paper, we focused on the properties of the PULX, in particular their spin-up rates, to estimate the surface magnetic dipole field strength $B$, beaming fraction $b$, and fastness parameter $\omega_*$ of the neutron star in each source.

For the PULX, we inferred the ranges for $B$, $\omega_*$, and $b$ based on different possibilities regarding the spin and luminosity states of the neutron star. We considered the alternative cases: (i) the observed spin-up rates along with an efficient standard spin-up torque, assuming that the systems are away from spin equilibrium, (ii) the spin-equilibrium condition according to which the fastness parameter has almost reached its critical value such that the total torque acting on the neutron star is sufficiently close to zero, (iii) the critical luminosity condition according to which the current X-ray luminosity of the system is well represented by its maximum critical value $(L_{\rm X}=L_{\rm c})$, (iv) the observed spin-up rates result from the standard spin-up torque under the critical luminosity condition, (v) the spin-equilibrium and critical luminosity conditions both apply, and (vi) the subcritical luminosity condition $(L_{\rm X}<L_{\rm c})$ with either the observed spin-up rates or the spin-equilibrium condition.

\begin{deluxetable*}{lccc}
\tablecolumns{4} \tablewidth{0pt} \tablecaption{Magnetic field, beaming fraction, and fastness parameter inferred from spin-up rate and critical luminosity. \label{tab:sucrlum}}
\tablehead{\colhead{Source}\hspace{2.85cm} & \colhead{$b$\hspace{1.85cm}} & \colhead{$\omega_*$\hspace{1.85cm}} & \colhead{$B\left(\times 10^{12}\,{\rm G}\right)$}} \startdata
ULX NGC 5907 & \hspace{1.85cm}0.19--0.30\hspace{1.85cm} & \hspace{1.85cm}0.16--0.61\hspace{1.85cm} & 27--38 \\
ULX NGC 7793 P13 & \hspace{1.85cm}0.021--0.027\hspace{1.85cm} & \hspace{1.85cm}0.065--0.14\hspace{1.85cm} & 0.12--0.32 \\
M82 X-2 (J095551$+$6940.8) & \hspace{1.85cm}0.16--0.25\hspace{1.85cm} & \hspace{1.85cm}0.060--0.22\hspace{1.85cm} & 2.0--2.9 \\
NGC 300 ULX1 & \hspace{1.85cm}$\sim 1$\hspace{1.85cm} & \hspace{1.85cm}0.015\hspace{1.85cm} & 6.8 \\
M51 ULX-7 & \hspace{1.85cm}0.052--0.077\hspace{1.85cm} & \hspace{1.85cm}0.019--0.063\hspace{1.85cm} & 0.62--1.0 \\
Swift J0243.6+6124 & \hspace{1.85cm}$\sim 1$\hspace{1.85cm} & \hspace{1.85cm}0.041\hspace{1.85cm} & 3.5 \\
\enddata
\tablecomments{Here, the magnetic field, beaming fraction, and fastness parameter of Swift~J0243.6+6124 are marginally obtained at\\$b\simeq1$ for $M_*=1.2\,M_{\odot}$ and $R_*=10\,{\rm km}$ and no common solution is found for NGC~1313~X-2 (see Section~\ref{sucr}). For all\\other sources, we use $M_*=1.4M_\odot$ and $R_*=10\,{\rm km}$.}
\end{deluxetable*}

The PULX are presumably accreting systems with $\omega_*<1$. Considering the observed spin-up rates alone, the upper limit of the fastness parameter $(\omega_*=1)$ can be used to estimate the lower limit of $b$ and the strong magnetic field range, whereas the upper limit of the beaming fraction, $b\leq 1$, can be used to obtain the weak magnetic field range, taking into account the lower limit of the fastness parameter, i.e., $\omega_*>(R_*/R_{\rm co})^{3/2}$.
The limits of the magnetic field range for each source also depend on the relative width of the boundary region in accordance with the theory of magnetosphere-disk interaction for ULX \citep[see also Section~\ref{model}]{eea+19}. Considering the spin-equilibrium condition alone, for which $\omega_*=\omega_{\rm c}=0.75$, we refer to the condition of the super-Eddington rate of mass transfer $(\dot{M}_0>\dot{M}_{\rm E})$ and obtain the lower limit of $b$ corresponding to the weak magnetic field range. The upper limit of the beaming fraction, which is $b=1$ in this case, determines the strong magnetic field range. Considering the critical luminosity
condition alone, the highest and lowest values of the minimum magnetic field are given by the upper and lower limits of the beaming fraction. The upper limit of the beaming fraction does not always extend to $b=1$. The accretion condition for a slow rotator, i.e., $\omega_*<1$, must be satisfied, which, in some cases, sets a maximum value of $b<1$ (see, e.g., the case of ULX~NGC~7793~P13 in Section~\ref{critB}). The lower limit of $b$ does not have a unique value either. Indeed, the lower limit of the fastness parameter for different values of the boundary region width yields the range for the lower limit of $b$ (Section~\ref{critB}).

We obtain tighter constraints on the magnetic fields, beaming fractions, and fastness parameters of the PULX using either the observed spin-up rates or the spin-equilibrium condition along with the critical luminosity condition. The intersection of the model curves delineating our magnetic field inference from the spin-up rate and critical luminosity reveals the range for the minimum magnetic dipole field strength on the neutron-star surface (\autoref{tab:sucrlum}). The minimum $B$ values inferred from both the observed spin-up rates and critical luminosity condition (\autoref{tab:sucrlum}) are well above the weak field range inferred from the spin-up rates alone (\autoref{tab:spinup}). Both the spin equilibrium and critical luminosity conditions can apply simultaneously in ULX~NGC~5907 and ULX~NGC~7793~P13 (\autoref{tab:eqcrlum}). In M82~X-2 and NGC~300~ULX1 (\autoref{Bwfromlum}), on the other hand, the range of $\omega_*$ inferred from the critical luminosity assumption remains well below the critical value of the fastness parameter $(\omega_{\rm c}=0.75)$, implying the latter two sources cannot be X-ray luminous at the maximum critical value while being close to spin equilibrium.

We also study the PULX assuming subcritical X-ray luminosities as the most general case for inferring the ranges of the surface dipole field strengths, beaming fractions, and fastness parameters either from the observed spin-up rates (\autoref{tab:susclum}) or from the spin-equilibrium condition (\autoref{tab:eqsclum}). When the spin-up rates are used, the upper limit of the fastness parameter $(\omega_*=1)$, corresponding to the lower limit of $b$, determines the upper limit of the $B$ range. The upper limit of $b$, on the other hand, follows from the critical luminosity condition and determines the lower limits of $B$ and $\omega_*$. When the spin-equilibrium condition is employed for $\omega_*=\omega_{\rm c}=0.75$, the lower and upper limits of the weak and strong magnetic field ranges, respectively, inferred from the spin-equilibrium condition alone, determine the $B$ range. The limits of the range for the beaming fraction remain also unaffected by the subcritical luminosity condition provided the PULX are close enough to spin equilibrium.

For the ultra-luminous X-ray pulsars, such as ULX~NGC~5907, M82~X-2, and NGC~300~ULX1, the ranges for the magnetic field strengths and beaming fractions inferred from the observed spin-up rates and subcritical luminosity condition partially overlap to yield the common range for ULX~NGC~5907 and M82~X-2 as $B\simeq (2.7-5.0)\times10^{13} \,{\rm G}$ for $b\simeq 0.16-0.25$ and the common range for ULX~NGC~5907 and NGC~300~ULX1 as $B\simeq (2.7-25)\times10^{13} \,{\rm G}$ for $b\simeq 0.25-0.30$. The range of overlap for NGC~300~ULX1, Swift~J0243.6+6124, and NGC~1313~X-2, is $B\simeq (4.0-25)\times10^{13} \,{\rm G}$ and $b\simeq 0.97-1$. The partial overlap of the individual ranges in ULX~NGC~7793~P13 and M51~ULX-7 yields $B\simeq (0.62-5.0)\times10^{12} \,{\rm G}$ and $b\simeq 0.020-0.027$. The source ULX~NGC~7793~P13, however, differs from other PULX by an order of magnitude in both the magnetic field strength and beaming fraction. Both $B$ and $b$, inferred from the spin-up rate and subcritical luminosity, are much smaller in ULX~NGC~7793~P13 compared to other PULX (\autoref{tab:susclum}).

Based on the spin-equilibrium and subcritical luminosity conditions, the ranges for $B$ and $b$ in ULX~NGC~5907 partially overlap those in NGC~300~ULX1, M82~X-2, ULX~NGC~7793~P13, Swift~J0243.6+6124, NGC~1313~X-2, and M51~ULX-7 to approximately determine the common ranges for the magnetic field and beaming fraction as $B\simeq (3.2-21)\times10^{13} \,{\rm G}$ and $b\simeq 0.1-1$, respectively (\autoref{tab:eqsclum}).

\subsection{Comparison With Observations}

We applied all plausible scenarios to each source in the current sample of PULX. The sources in the sample, however, differ from each other in some of their properties such as the orbital period of the binary, the type of the donor star, and the spin state of the neutron

\begin{deluxetable*}{lccc}
\tablecolumns{4} \tablewidth{0pt} \tablecaption{Magnetic field, beaming fraction, and fastness parameter inferred from spin equilibrium and critical luminosity. \label{tab:eqcrlum}}
\tablehead{\colhead{Source}\hspace{2.35cm} & \colhead{$b$\hspace{2.35cm}} & \colhead{$\omega_*=\omega_{\rm c}$\hspace{2.35cm}} & \colhead{$B\left(\times 10^{13}\,{\rm G}\right)$}} \startdata
ULX NGC 5907 & \hspace{2.35cm}0.50--1\hspace{2.35cm} & \hspace{2.35cm}0.75\hspace{2.35cm} & 5.6--9.4 \\
ULX NGC 7793 P13 & \hspace{2.35cm}0.12--1\hspace{2.35cm} & \hspace{2.35cm}0.75\hspace{2.35cm} & 0.22--1.2 \\
\enddata
\tablecomments{Here, we use $M_*=1.4M_\odot$ and $R_*=10\,{\rm km}$ for both PULX. For all other sources, we found no solution even for a\\small value of the neutron-star mass, i.e., for $M_*=0.9M_\odot$ when $R_*=10\,{\rm km}$ (see Section~\ref{eqcr} for the effect of larger $R_*$\\used together with sufficiently small $M_*$ on $B$).}
\end{deluxetable*}

\noindent star. For each individual source we expect some of the scenarios to be more likely than the others.

The optical spectrum of NGC~7793~P13 indicates a blue supergiant companion while the photometric analysis reveals a long orbital period of more than 2 months \citep{motch+14,fabrika+15,furst+18}. ULX~NGC~7793~P13 exhibits a slow but persistent spin up \citep{fue+16}.
NGC~300~ULX1 is another PULX with a supergiant companion. The photospheric absorption lines in the near-infrared spectrum of NGC~300~ULX1 allowed \citet{hei+19} to identify the donor star as a red supergiant, which, assuming Roche lobe overflow, indicates an orbital period of $\sim1-2$ years. A long-term constant spin-up rate with a time span of $\sim1$ year is probably due to the small fastness parameter of NGC~300~ULX1 if the system is away from spin equilibrium \citep{vas+19}. According to our field inference from spin-up rates and subcritical luminosity, both sources might be slow rotators with $\omega_*\gtrsim0.065$ for ULX~NGC~7793~P13 and $\omega_*\gtrsim0.015$ for NGC~300~ULX1 (see \autoref{tab:susclum}). It is also possible that ULX~NGC~7793~P13 is close to spin equilibrium with $\omega_*\simeq 0.75$ as the source exhibits slow spin-up of rate $\sim -10^{-12}\,{\rm s\,s^{-1}}$ (\autoref{tab:eq}). We infer $B\sim 10^{11}-10^{13}\,{\rm G}$ for ULX~NGC~7793~P13 whether or not the source is close to spin equilibrium and $B\sim 10^{13}-10^{15}\,{\rm G}$ for NGC~300~ULX1.

\begin{figure*}
    \centering
            \includegraphics[width=0.85\textwidth]{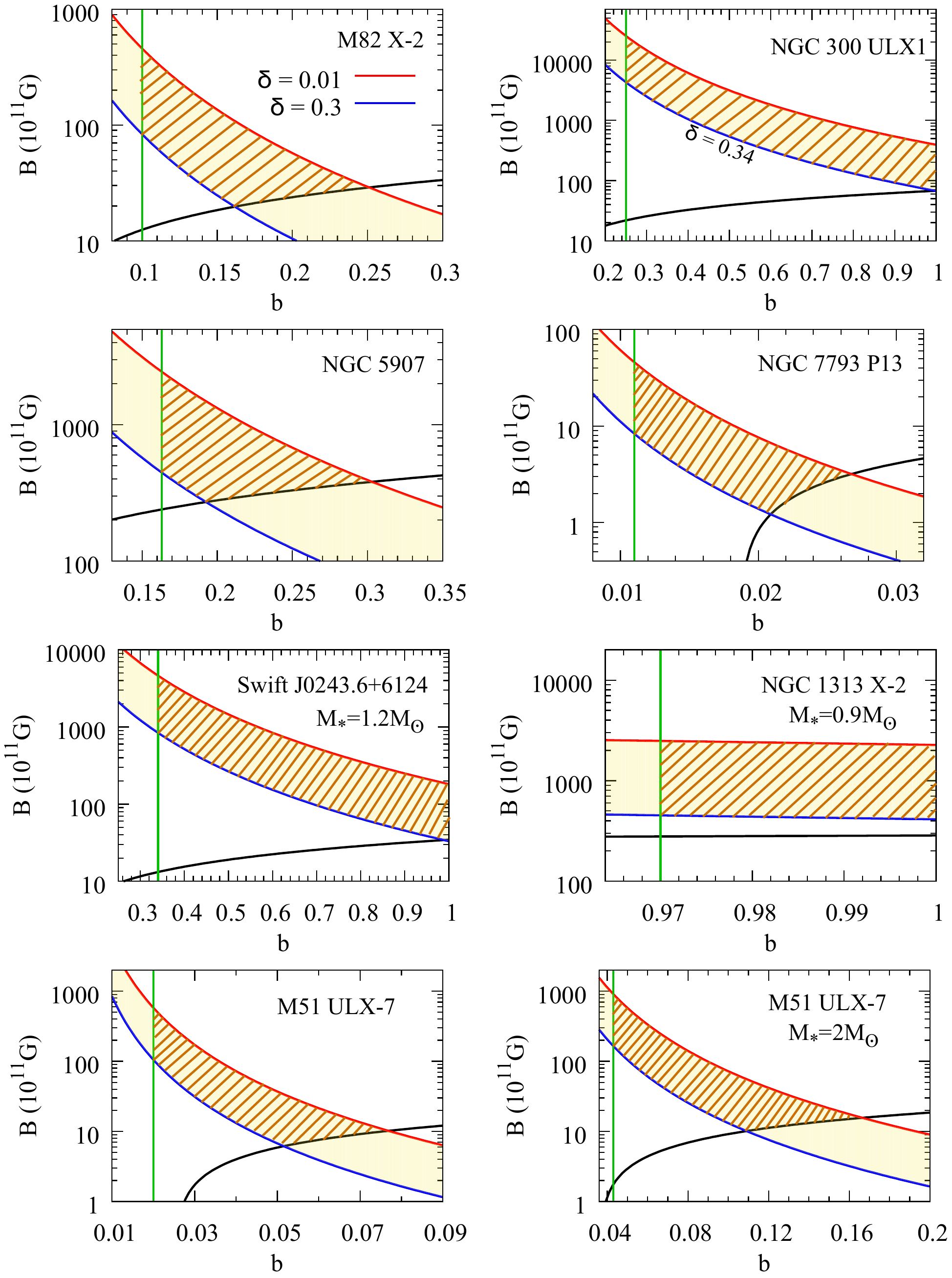}             
    \caption{Estimation of the magnetic field (striped region) in terms of the beaming fraction using the spin-up rates of PULX at subcritical luminosity. The striped region lies within the shaded region between the red and blue solid curves ($B$ inferred from spin-up rates) and is bounded from left by the vertical green solid line (lower limit of beaming given by $\omega_*=1$) and from below by the black solid curve ($B$ inferred from critical luminosity) if the black curve intersects with the shaded region. Unless otherwise stated (e.g. NGC~1313~X-2 and Swift~J0243.6+6124), we used $M_*=1.4M_\odot$ and $R_*=10\,{\rm km}$ for all sources. The bottom left and right panels for M51~ULX-7 compare two different cases with $M_*=1.4M_\odot$ and $M_*=2M_\odot$, respectively.}
   \label{Bsusub}
\end{figure*}

\begin{figure*}
    \centering
            \includegraphics[width=0.85\textwidth]{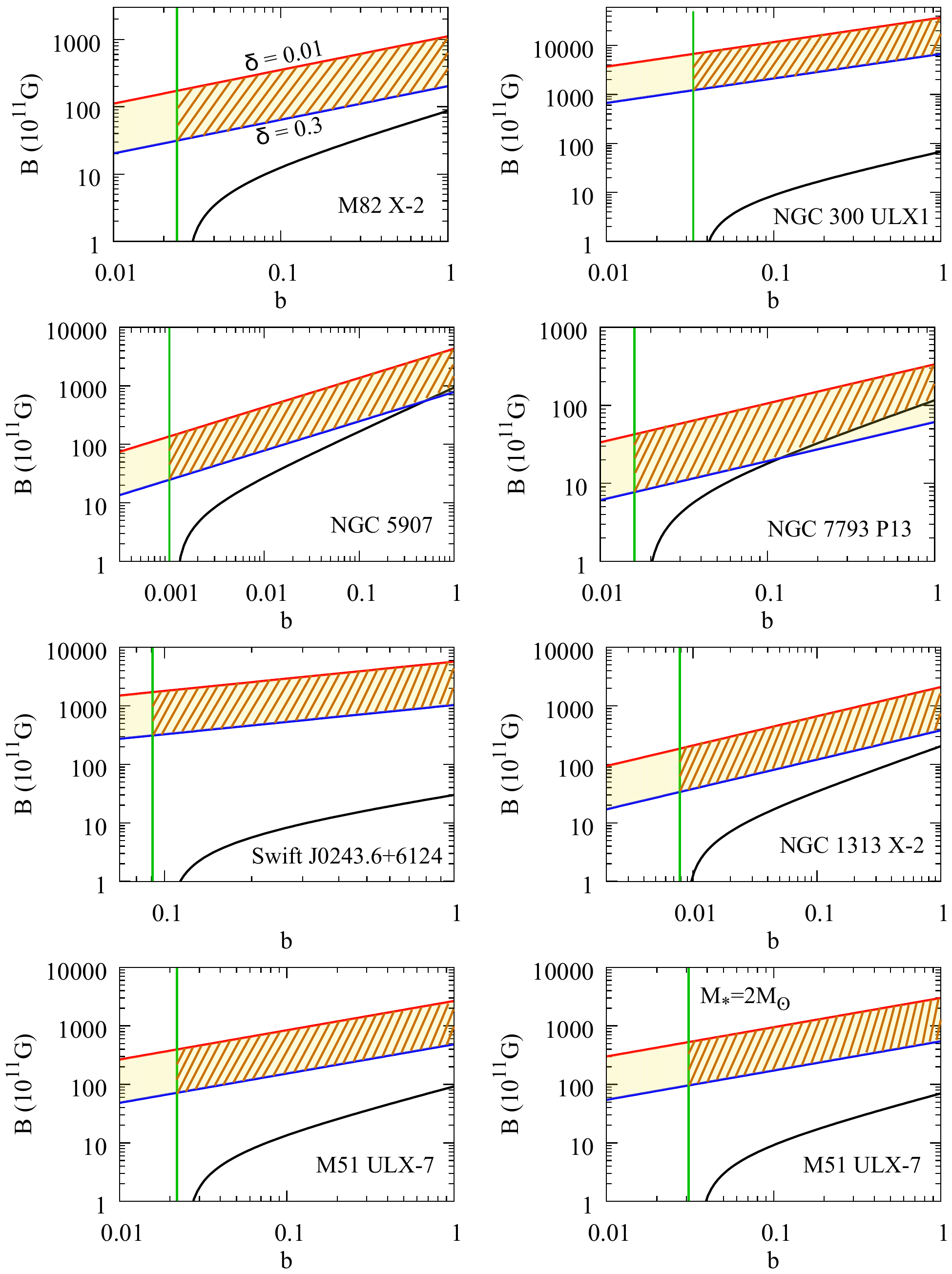}             
    \caption{Estimation of the magnetic field (striped region) in terms of the beaming fraction using the spin equilibrium of PULX at subcritical luminosity. The striped region lies within the shaded region between the red and blue solid curves ($B$ inferred from spin-equilibrium) and is bounded from left by the vertical green solid line (lower limit of beaming given by \autoref{lwlimb}) and from below by the black solid curve ($B$ inferred from critical luminosity) if the black curve intersects with the shaded region. We used $M_*=1.4M_\odot$ and $R_*=10\,{\rm km}$ for all sources. The bottom left and right panels for M51~ULX-7 compare two different cases with $M_*=1.4M_\odot$ and $M_*=2M_\odot$, respectively.}
   \label{Beqsub}
\end{figure*}

The intermittent pulsations and large variations of the spin-up rate together with a secular spin-down epoch observed in M82~X-2 indicate that this PULX might be close to spin equilibrium. It follows from timing that the orbital period is $\sim2.5\,{\rm d}$ and the mass of the companion in M82~X-2 is likely to be in the 5--20$\,M_\odot$ range, which is typical of a main sequence O-type star \citep{bac+20}. Another PULX with a short orbital period is ULX~NGC~5907 for which a secular spin-up rate and an orbital period of $\sim5\,{\rm d}$ were inferred \citep{isr+16a}. Using the spin-up rates at subcritical luminosity, we obtain similar values for the beaming fraction in the $\sim 0.1-0.3$ range for both sources (\autoref{tab:susclum}). Assuming that both sources are away from spin equilibrium, the neutron-star magnetic field we infer from spin-up rates is more than one order of magnitude higher in ULX~NGC~5907 than in M82~X-2 (\autoref{Bsusub}). If, however, M82~X-2 is close to spin equilibrium and ULX~NGC~5907 is not, the field inferred from spin equilibrium for M82~X-2 will be comparable to the $B$ values estimated through spin-up rates for ULX~NGC~5907. Using a beaming fraction $b\sim 0.25$ for both PULX, we find $B\sim (1-5)\times 10^{13}\,{\rm G}$ for M82~X-2 (\autoref{Beqsub}) and $B\sim (3-6)\times 10^{13}\,{\rm G}$ for ULX~NGC~5907 (\autoref{Bsusub}).
\vspace{0.06in}

One of the most recently discovered PULX in M51 with a secular spin up seems to have a short orbital period as well. A lower limit of $8\,M_\odot$ on the mass of the companion in M51~ULX-7, which is suspected to be an OB giant, was estimated using an orbital period of $\sim2\,{\rm d}$. The magnetic dipole field on the surface of the neutron star was proposed to be around $10^{12}-10^{13}\,{\rm G}$ for $0.08\lesssim b\lesssim 0.25$ \citep{rod+20}. Our results suggest $B\simeq (0.6-1)\times 10^{12}\,{\rm G}$ and $0.05\lesssim b\lesssim 0.08$ using spin-up rates and critical luminosity condition (\autoref{Bwsulum3}), whereas a larger field range of $B\simeq (0.6-60)\times 10^{12}\,{\rm G}$ is found with $0.02\lesssim b\lesssim 0.08$ using spin-up rates with subcritical luminosity (\autoref{tab:susclum}). The magnetic field lies between $\sim 7\times 10^{12}\,{\rm G}$ and $\sim 3\times 10^{14}\,{\rm G}$ for a beaming fraction in the $\sim 0.02-1$ range if the system is close to spin equilibrium with a subcritical luminosity (\autoref{tab:eqsclum}). Our field values based on spin-equilibrium assumption are consistent with the field estimates of \citet{vas+20}. Note that these ranges for $B$ can have lower values for relatively large radii and small masses for the neutron star in M51~ULX-7 (Sections~\ref{Bfrmeq} and \ref{eqcr}).

Although not well-established, an orbital period of $\lesssim2\,{\rm d}$ and a B-type companion with an upper mass limit of $12\,M_\odot$ were suggested for another recently discovered PULX in NGC~1313. Among other PULX, ULX~NGC~1313~X-2 seems to be quite unusual as there appears to be a secular spin down over 100 days between the two detections of pulses. During both pulse \,\,detection \,\,\,epochs, \,\,\,however, \,\,the \,\,measured \,\,frequency

\begin{deluxetable*}{lccc}
\tablecolumns{4} \tablewidth{0pt} \tablecaption{Magnetic field, beaming fraction, and fastness parameter inferred from spin-up rate and subcritical luminosity. \label{tab:susclum}}
\tablehead{\colhead{Source}\hspace{2.8cm} & \colhead{$b$\hspace{2cm}} & \colhead{$\omega_*$\hspace{2cm}} & \colhead{$B\left(\times 10^{13}\,{\rm G}\right)$}} \startdata
ULX NGC 5907 & \hspace{2cm}0.16--0.30\hspace{2cm} & \hspace{2cm}0.16--1\hspace{2cm} & 2.7--25 \\
ULX NGC 7793 P13 & \hspace{2cm}0.011--0.027\hspace{2cm} & \hspace{2cm}0.065--1\hspace{2cm} & 0.012--0.50 \\
M82 X-2 (J095551$+$6940.8) & \hspace{2cm}0.10--0.25\hspace{2cm} & \hspace{2cm}0.060--1\hspace{2cm} & 0.20--5.0 \\
NGC 300 ULX1 & \hspace{2cm}0.25--1\hspace{2cm} & \hspace{2cm}0.015--1\hspace{2cm} & 0.68--240 \\
M51 ULX-7 & \hspace{2cm}0.020--0.077\hspace{2cm} & \hspace{2cm}0.019--1\hspace{2cm} & 0.062--6.0 \\
NGC 1313 X-2 & \hspace{2cm}0.97--1\hspace{2cm} & \hspace{2cm}0.91--1\hspace{2cm} & 4.0--25 \\
Swift J0243.6+6124 & \hspace{2cm}0.34--1\hspace{2cm} & \hspace{2cm}0.041--1\hspace{2cm} & 0.35--46 \\
\enddata
\tablecomments{Here, the parameter ranges of NGC~1313~X-2 and Swift~J0243.6+6124 are obtained for $M_*=0.9\,M_{\odot}$ and\\$M_*=1.2\,M_{\odot}$, respectively, when $R_*=10\,{\rm km}$. For all other sources, we use $M_*=1.4M_\odot$ and $R_*=10\,{\rm km}$.}
\end{deluxetable*}

\noindent derivatives indicate spin up rather than spin down \citep{sat+19}. As also suggested by observations, ULX~NGC~1313~X-2 might be close to spin equilibrium and might therefore have $B\sim 10^{12}-10^{14}\,{\rm G}$ with a minimum beaming fraction of $\sim 0.01$ (see \autoref{Beqsub} and \autoref{tab:eqsclum}).
\vspace{0.1in}

The first galactic PULX has been associated with a newly discovered Be/X-ray binary Swift~J0243.6+6124. The system is known as an X-ray transient accreting from a Be-type main sequence donor with a mass of $16\,M_\odot$ and has an orbital period of $\sim28\,{\rm d}$. This PULX spins up in correlation with the X-ray flux. The higher the X-ray flux gets, the more dramatic is the neutron-star spin-up rate \citep{wil+18}. The system is an X-ray transient and can therefore be either close to or away from spin equilibrium. A sudden increase in the mass accretion rate can push the ULX out of spin equilibrium causing the source to spin up with observed rates from which we deduce $B\sim 10^{12}-10^{14}\,{\rm G}$, $b> 0.3$, and $\omega_*\gtrsim 0.04$ at subcritical luminosities (\autoref{tab:susclum}). The source may also be close to spin equilibrium with $\omega_*\simeq 0.75$ and $B\sim 10^{13}-10^{14}\,{\rm G}$ for $b\gtrsim 0.1$ (\autoref{tab:eqsclum}).

\subsection{Field Strength, Beaming, and Spin State}

Whether or not the sources are away from spin equilibrium, the neutron star in NGC~300~ULX1 seems to possess the strongest magnetic dipole field on its surface within the current family of PULX. It is highly likely, on the other hand, that the weakest magnetic dipole field is on the surface of the neutron star in ULX~NGC~7793~P13. In general, the surface magnetic dipole field strength of all PULX covers the $\sim 10^{11}-10^{15}\,{\rm G}$ range (\autoref{tab:susclum} and \autoref{tab:eqsclum}). Our analysis indicates that the PULX are neutron stars accreting matter from super-Eddington disks with a range of dipole magnetic fields depending on the circumstances. The field range inferred is typical of young neutron stars, like the young pulsar population; possible fields extend to but are not necessarily restricted to magnetar strengths.

The present day field strengths we infer from the observed spin-up rates or the spin-equilibrium condition, may, however, correspond to the already decayed fields of initial strengths in the magnetar range if magnetar strength dipole fields actually exist in newborn neutron stars and decay with time. Such a possibility is also in line with the recent scenario proposed by \citet{eea+19} to describe how the very early phase of neutron-star evolution in HMXBs can spawn seemingly different classes of ULX. According to \citet{eea+19}, the ULX with initial magnetic dipole field strengths $\gtrsim 10^{13}\,{\rm G}$ can evolve into PULX when the neutron-star spin period becomes sufficiently long (e.g., $P\gtrsim 1\,{\rm s}$). In particular, the neutron stars with initial fields of magnetar strengths, e.g., $B(t=0)=10^{15}\,{\rm G}$, can spin down to periods in excess of several tens of seconds for relatively wide zones $(\delta>0.1)$ of magnetosphere-disk interaction \citep{eea+19}. The present day field strength we infer for NGC~300~ULX1 lies in the $\sim 10^{13}-10^{15}\,{\rm G}$ range (\autoref{tab:susclum} and \autoref{tab:eqsclum}). As suggested by the current results we obtain here, NGC~300~ULX1 with a spin period of $\sim 30\,{\rm s}$ (\autoref{tab:eq}) might be an indicator for the presence of super strong initial magnetic dipole fields on the surfaces of some newborn neutron stars in HMXBs.

The ULX are usually thought to be young, disk accreting systems fed by either wind or Roche lobe overflow from the normal star. They may never achieve true spin equilibrium, instead they might alternate between accretion \,and \,\,propeller \,\,states, \,when \,the \,\,neutron \,\,star

\begin{deluxetable*}{lccc}
\tablecolumns{4} \tablewidth{0pt} \tablecaption{Magnetic field, beaming fraction, and fastness parameter inferred from spin equilibrium and subcritical luminosity. \label{tab:eqsclum}}
\tablehead{\colhead{Source}\hspace{2.85cm} & \colhead{$b$\hspace{2.25cm}} & \colhead{$\omega_*=\omega_{\rm c}$\hspace{2.25cm}} & \colhead{$B\left(\times 10^{13}\,{\rm G}\right)$}} \startdata
ULX NGC 5907 & \hspace{2.25cm}0.001--1\hspace{2.25cm} & \hspace{2.25cm}0.75\hspace{2.25cm} & 0.24--45 \\
ULX NGC 7793 P13 & \hspace{2.25cm}0.016--1\hspace{2.25cm} & \hspace{2.25cm}0.75\hspace{2.25cm} & 0.077--3.3 \\
M82 X-2 (J095551$+$6940.8) & \hspace{2.25cm}0.024--1\hspace{2.25cm} & \hspace{2.25cm}0.75\hspace{2.25cm} & 0.30--10 \\
NGC 300 ULX1 & \hspace{2.25cm}0.033--1\hspace{2.25cm} & \hspace{2.25cm}0.75\hspace{2.25cm} & 12--350 \\
M51 ULX-7 & \hspace{2.25cm}0.022--1\hspace{2.25cm} & \hspace{2.25cm}0.75\hspace{2.25cm} & 0.70--26 \\
NGC 1313 X-2 & \hspace{2.25cm}0.0078--1\hspace{2.25cm} & \hspace{2.25cm}0.75\hspace{2.25cm} & 0.35--21 \\
Swift J0243.6+6124 & \hspace{2.25cm}0.091--1\hspace{2.25cm} & \hspace{2.25cm}0.75\hspace{2.25cm} & 3.2--57 \\
\enddata
\tablecomments{For all sources, we use $M_*=1.4M_\odot$ and $R_*=10\,{\rm km}$.}
\end{deluxetable*}

\noindent spins up and down, respectively. The presumption of spin equilibrium can be as reliable as the use of measured spin-up rates to infer magnetic field. The PULX might be in the course of spin up while being close to spin equilibrium. In this paper, we refer to the state of near spin equilibrium keeping $\omega_*=\omega_{\rm c}=0.75$ \citep{tur+17} and the state of non-equilibrium employing an efficient standard spin-up torque. Our analysis is not based on a specific torque model with a particular dependence on the fastness parameter. We rather focus on either the spin-equilibrium or spin-up state in which the torque acting on the neutron star is well established.
\vspace{0.1in}

The question naturally arises why we observe X-ray pulsations from some ULX but not from others. The elusiveness of pulsations was attributed by \citet{eks+15} to the presence of an optically thick medium due to the stellar wind of the companion. A more detailed account is given by \citet{mus+17}. Our calculations here suggest that beaming may also play a role for understanding the lack of pulsations from some of the neutron-star ULX. In the present work, we come up with a new approach to the beaming fraction. The beaming fraction depends on both the mass accretion rate and magnetic field strength as well as several other parameters. Our definition can be specifically reliable for pulsating sources. The results of our analysis are, however, independent of the specific definition of beaming fraction and therefore have robust implications within the frame of our model (Section~\ref{sec:beam}).
\vspace{0.1in}

Strong gravitational field effects on the emerging radiative flux near the surface of the neutron star are not included in our beaming definition. The relativistic effects such as light bending can smooth out the pulse profile of a pulsar to some degree. However, the mass accretion rates and luminosities in PULX are so high that the accretion curtain on the magnetospheric surface can behave as an optically thick shield around the neutron star to give the highest contribution to the smoothing of the pulse profile while rendering the relativistic effects undetectable \citep{mus+17}. The accretion column can be as high as the neutron-star radius for sufficiently high mass accretion rates. The direct component of beamed X-ray flux from the column dominates over the reflected component from the neutron-star surface \citep{mus+18}. In the presence of an optically thick envelope around the neutron star, the direct component can be emitted out of the accretion column at radial distances more than twice the neutron-star radius.
The effect of light bending and redshift on the pulsar beam characteristics is strongest when emission is close to the neutron-star surface. The flattening of the pulse profile is strongest for neutron stars with smallest radii, but becomes insignificant for emission regions beyond 20 km \citep{kap91}. We therefore expect strong gravitational field effects on the pulse profile of PULX to be negligible as long as sufficiently high mass-accretion rates and strong magnetic fields are concerned.
\vspace{0.1in}

It was suggested \citep{eks+15} that the reduction of the scattering cross-section \citep{can71,pac92} in super-strong magnetic fields, a mechanism employed for the super-Eddington outbursts of isolated magnetars in our galaxy, may be at work in M82~X-2 (NuSTAR~J095551$+$6940.8) \citep[see also][]{ton15a,ton15b,dal+15}. This argument relies on the assumption that the object is near spin equilibrium so that indeed very strong magnetic fields are required to explain the spin-up rate of the object. If the system is in outburst and so out of spin equilibrium then the observed spin-up rate does not require super-critical magnetic fields, thus favoring the beaming argument.

\section{Conclusions} \label{conc}

We found different solutions for the surface magnetic dipole fields $B$, beaming fractions $b$, and fastness parameters $\omega_*$ of the neutron stars in the PULX M82~X-2, NGC~300~ULX1, ULX~NGC~5907, ULX~NGC~7793~P13, M51~ULX-7, NGC~1313~X-2, and Swift~J0243.6+6124 with the help of alternative assumptions either used alone or combined with each other. These assumptions correspond to different possibilities of the neutron-star spin and luminosity states.

We considered the following alternative scenarios: (i) the PULX are away from spin equilibrium; an efficient standard spin-up torque can be used to account for the observed spin-up rates, (ii) the PULX are so close to spin equilibrium that the fastness parameter is given by its critical value, (iii) the X-ray luminosities of the PULX can be well represented by the maximum critical luminosity, (iv) the conditions described in (i) and (iii) both apply, (v) the conditions described in (ii) and (iii) both apply, and (vi) the X-ray luminosities of the PULX are subcritical and either the condition described in (i) or the condition described in (ii) applies.

We found the narrowest ranges for $B$, $b$, and $\omega_*$ when we used the critical luminosity condition along with either the observed spin-up rates or the spin-equilibrium condition. The scenario (iv) based on the observed spin-up rates at critical luminosity worked well for all the PULX yielding $B$ in the $\sim 10^{11}-10^{13}\,{\rm G}$ range. The scenario (v) based on the spin-equilibrium condition at critical luminosity, on the other hand, worked only for the two PULX yielding $B$ in the $\sim 10^{12}-10^{14}\,{\rm G}$ range. We obtained wider ranges for $B$, $b$, and $\omega_*$ when we assumed subcritical luminosities along with either the observed spin-up rates or the spin-equilibrium condition. In the subcritical luminosity regime, using both the observed spin-up rates and spin-equilibrium condition, we found $B$ in the $\sim 10^{11}-10^{15}\,{\rm G}$ range covering the individual ranges for the magnetic field of each PULX. The individual ranges for the magnetic fields and beaming fractions inferred from the observed spin-up rates and subcritical luminosity condition partially overlap at the common range $B\simeq (2.7-5.0)\times10^{13} \,{\rm G}$ and $b\simeq 0.16-0.25$ for ULX~NGC~5907 and M82~X-2. For ULX~NGC~5907 and NGC~300~ULX1, the range of overlap is $B\simeq (2.7-25)\times10^{13} \,{\rm G}$ and $b\simeq 0.25-0.30$. For NGC~300~ULX1, Swift~J0243.6+6124, and NGC~1313~X-2, we find the overlap range as $B\simeq (4.0-25)\times10^{13} \,{\rm G}$ and $b\simeq 0.97-1$. The individual ranges in ULX~NGC~7793~P13 and M51~ULX-7 partially overlap at $B\simeq (0.62-5.0)\times10^{12} \,{\rm G}$ and $b\simeq 0.020-0.027$. The ranges for $B$ and $b$ inferred from the spin-equilibrium and subcritical luminosity conditions in ULX~NGC~5907 partially overlap those in NGC~300~ULX1, M82~X-2, ULX~NGC~7793~P13, Swift~J0243.6+6124, NGC~1313~X-2, and M51~ULX-7 to determine the common ranges $B\simeq (3.2-21)\times10^{13} \,{\rm G}$ and $b\simeq 0.1-1$. We conclude that the surface magnetic dipole fields of neutron stars in PULX are not necessarily of magnetar strength. The $B$ ranges we estimated through different scenarios also include values well below the quantum critical limit for the magnetic field. Weaker (conventional, non-magnetar) $B$ fields are inferred if the luminosity is taken to be critical. Observations implying critical luminosity conditions will thus narrow down the inferred field ranges.

Our results suggest that the role of beaming cannot be neglected. Subcritical luminosities can still be realized for $b<1$ even in the absence of magnetar-strength fields, which would otherwise increase the critical luminosity by reducing the scattering cross-section. The lack of pulsations from the ULX that contain neutron stars can also be understood in terms of beaming. We would roughly expect to have a factor of $1/b$ more neutron-star ULX than the current number of PULX. For $b=0.01$ on average, we would have $\sim 700$ non-pulsating neutron-star ULX if the number of PULX was $\sim 7$. The ULX with neutron stars would then correspond to a significant fraction of the total number of ULX.

\section*{Acknowledgements}
We thank the referee for constructive comments and suggestions. KYE and MT acknowledge support from the Scientific and Technological Research Council of Turkey (T\"{U}B\.{I}TAK) with the project number 112T105. MAA is a member of the Science Academy (Bilim Akademisi), Turkey.

\bibliographystyle{aasjournal}
\footnotesize{
\bibliography{refs} 
}

\end{document}